\documentclass[12pt]{iopart}

\usepackage{blindtext}
\usepackage{amssymb}
\usepackage{lipsum}
\usepackage{braket}
\usepackage{graphicx}
\usepackage{dcolumn}
\usepackage{bm}
\usepackage{subfigure}
\usepackage[normalem]{ulem}
\usepackage{amssymb}
\usepackage{lipsum}

\begin{document}

\title[Non-commutative foliated quantum gravity and the wave function of the universe]{Cosmic inflation in an extended non-commutative foliated quantum gravity: the wave function of the universe}

\author{
C\'esar A. Zen Vasconcellos$^{1,2}$, 
Peter O. Hess$^{3,4}$,
Jos\'e de Freitas Pacheco$^{5}$,
Fridolin Weber$^{6,7}$,
Benno Bodmann$^{1}$,
Dimiter Hadjimichef$^{1}$,
Geovane Naysinger$^{1}$,
Rodrigo Fraga$^{1}$, Jo\~ao G.G. Gimenez$^{1}$,
Marcelo Netz-Marzola$^{4}$,
Mois\'es Razeira$^{8}$
}

\address{
$^{1}$ Instituto de Fisica, Universidade Federal do Rio Grande do 
Sul (UFRGS), Porto Alegre, Brazil\\
$^{2}$ International Center for Relativistic Astrophysics 
Network (ICRANet), Pescara, Roma, Italy\\
$^{3}$ Instituto de Ciencias Nucleares, Universidad Nacional Autonoma de Mexico (UNAM), A.P. 70-543, Mexico City, Mexico\\
$^{4}$ Frankfurt Institute for Advanced Studies (FIAS), J.W. von Goethe Universit\"at, Frankfurt am Main, Germany\\
$^{5}$ Observatoire de la Cote d'Azur (OCA), Nice, France\\
$^{6}$ Department of Physics, San Diego State University (SDSU), 
San Diego, USA\\
$^{7}$Department of Physics, University of California at San Diego (UCSD), La Jolla, USA\\
$^{8}$ Universidade Federal do Pampa (UNIPAMPA), Caçapava do Sul, Brazil
}

\begin{abstract}
We propose a novel extension to the recently developed non-commutative 
Riemannian foliated branch-cut quantum gravity (BCQG). Based on an extended Faddeev–Jackiw symplectic deformation of the conventional Poisson algebra, we investigate non-commutativity effects on a symplectic topological manifold that provides a natural isomorphic setting composed by a triad of canonically conjugate 
scalar complex fields which comprise quantum complementary dualities. 
The extended manifold developed here includes the canonical BCQG
cosmic scale factor and its quantum counterparts. It integrates
concepts from Hermann Weyl’s perfect fluid domain and introduces a
scalar-complex field inspired by inflaton models.
This extension allows the elaboration of a scenario where the branch-cut domain coexists with a scalar-complex inflaton-type field, settled on a plateau of a strong-gravity interacting structure, shaped by the primordial energy, radiation-, quintessence-, stiff-, baryon- and dark-matter, and spacetime curvature  
which would drive cosmic inflation. The resulting symplectic algebraic extension allows additionally a liaison with primordial cosmic density perturbation predictions based on heuristic derivations of the scale invariance of the primordial spectrum of the universe, as well as with eternal inflation and primordial chaotic cosmic mechanisms proposals. Given the peculiar characteristic of BCQG in overcoming the primordial singularity, the non-commutative symplectic extension proposal, due to the complex character of the constituent fields, additionally outlines the evolutionary process of the mirror universe that encompasses analytically continued complex conjugate equations of the Friedmann-type. This brings unique characteristics to the main scenario for the BCQG cosmic evolution, with a mirrored parallel evolutionary universe, adjacent to ours, nested in the structure of space and time. 
The evolutionary process of the mirror universe occurs in the negative sector of cosmological thermal time, and is characterized by a continuous cosmic contraction, with a systematic and continuous increase in temperature and a decrease in entropy before reaching the transition region to the current stage of the universe. In contrast, the  subsequent present evolutionary stage of cosmic expansion is characterized by a positive complex cosmological-time sector, accompanied by a systematic decrease in temperature and increase in entropy. 

Based on a complementary analytically continued Friedmann-type
equation, combined with a quantum approach based on the Ho\v{r}awa-Lifshitz quantum gravity, we describe the dynamic evolution of the universe’s wave function, unfolding unprecedented predictions for the cosmic evolution and inflation.
The non-commutative foliated quantum gravity approach offers a new
perspective on explaining the accelerated cosmic expansion of the
universe,
strongly suggesting that
non-commutative algebra induces the late accelerated growth of both
the universe’s wave function and the corresponding scale factor, along
with their quantum counterparts.
In contrast to the conventional inflationary model, where inflation requires a remarkably fine-tuned set of initial conditions in a patch of the universe, non-commutative foliated quantum gravity, analytically continued to the complex plane, captures short and long scales of spacetime, leading to an evolutionary cosmic dynamic through a topological reconfiguration of the primordial cosmic matter and energy content. This result introduces new speculative framework elements regarding the reconfiguration of matter and energy due to an underlying non-commutative spatio-temporal structure as a driver of spacetime cosmic acceleration.
\end{abstract}

%
%
%
%
%

\section{Introduction}\label{1}

The chronological evolution of the universe comprises different phases, of which the recombination and decoupling periods stand out, as well as a ‘primordial’ era characterized by inflation, followed by the dominant phases of radiation, matter and, currently, a dark energy dominated period, which is widely assumed to be the main cause of the accelerated cosmic expansion.

In this work, we focus mainly on the inflationary period, proposing a novel extended approach to the recently developed 
non-commutative Riemannian foliated
branch-cut quantum gravity (BCQG)~\cite{Zen2024}.

We investigate the effects of imposing an extended enhanced non-commutative symplectic deformation of the conventional Poisson algebra on a topological manifold that conveys a natural isomorphic setting of a canonically conjugate triad of dual vector spaces. The extended approach encompasses the cosmic scale factor of BCQG and complementary dual quantum counterparts, outlined in the perfect Hermann Weyl fluid domain and a scalar complex quantum field inflaton-inspired~\cite{Guth1981}. This scenario allows the coexistence of the quantum domain of branch-cut gravity with an inflationary mechanism that, although controversial for assuming an entire universe starting essentially from nothing, has had remarkable success in explaining relevant qualitative and quantitative properties of the universe~\cite{Guth1981,Guth2004}. Furthermore, based on this extended scenario, additional conceptual elements, such as density perturbations, grounded on heuristic derivation of the scale invariance of the primordial spectrum of the universe, as well as predictions of an eternal inflation and even primordial chaotic mechanisms associated with the origin of the universe, contribute to broadening the descriptive basis of the original BCQG formulation in order to incorporate elements of standard  inflation conceptions.

Moreover, in view of the peculiar characteristics of BCQG in overcoming the primordial singularity, this formulation makes it possible, - due to the complex character of the non-commutative symplectic extended composition -, to make predictions about the evolutionary scenarios of the mirror universe by means of complex-conjugated Friedmann-type BCQG equations. This brings unique features to the main BCQG scenario of cosmic evolution, with a parallel, mirror-like, evolutionary universe adjacent to ours, nested in the fabric of space and time, with its evolutionary process in the negative sector of cosmological thermal time corresponding to a contraction process followed by a continuous expansion in its positive temporal sector. These characteristics open up a range of opportunities involving a reverse process to cosmic inflation in the mirror sector. 

Returning to the aforementioned scenario, the assumed continuous contraction of the mirror universe is accompanied by an increase in temperature and a decrease in entropy before reaching the cosmic transition region. In the subsequent domain to the transition region, the continuous expansion of the universe is accompanied by a systematic decrease in temperature and an increase in entropy corresponding to the positive sector of the complex cosmological time. 
This formulation allows us to describe the dynamical evolution of the
universe’s wave function and the cosmic scale factor. We detail these
along with their complementary quantum dual counterparts, achieving
unprecedented results.

\section{BCQG}\label{BCQG}

The proposed theoretical approach follows the branch-cut quantum gravity (BCQG) framework~\cite{Zen2024,Bodmann2023a,Bodmann2023b}, based on the WheelerDeWitt~\cite{WdW} and the Ho\v{r}ava-Lifshitz~\cite{Horava} formulations, incorporating elements of symplectic geometry and non-commutative algebra~\cite{Zen2024} to investigate the interplay between small-scale quantum effects and large-scale cosmic evolution. BCQG provides moreover an alternative descriptive approach for cosmic evolution in comparison to conventional inflationary theories by building a quantum gravity environment through a fundamental restructuring of the conventional topological spacetime geometry rather than relying on specific initial conditions and {\it ad hoc} mechanisms. 

Additionally, as stressed before~\cite{Zen2024}, this approach offers a fresh perspective on the generation of relic gravitational waves and a possible solution to unresolved cosmological puzzles, such as the horizon and flatness problems.
As stressed before~\cite{Zen2024}, 
the realization of a non-commutative algebra structure captures both
short and long spatio-temporal scales. This drives the evolution of
the universe’s wave function and cosmic scale factor, and reconfigures
matter on small to intermediate scales. It also prompts the generation
of relic gravitational waves, a topic for future investigation.

The symplectic non-commutative BCQG approach, as previously emphasized~\cite{Zen2024}, was assembled by means of a deformation of the conventional Poisson algebra, and enhanced with a symplectic metric, based on the Faddeev-Jackiw two-fields quantization formulation~\cite{Faddeev1967,Faddeev1988} and extended in this contribution to a triad of dual and complementary quantum fields. Motivated by the compactification study of the properties of topological spaces, the non-commutative symplectic algebra is reflected in the validity of differential graded algebra with finite dimensional cohomology (compactification) and a rational Poincar\'e duality~\cite{Kontsevich,Silva}, which comprises harmonic foliations with minimal leaves on a Riemannian manifold~\cite{Kamber}, a crucial topic to embrace BCGQ. In other words, the Poisson brackets have the form $\{q_i,p_j\} = g_{ij}$, and because of the antisymmetry of the Poisson brackets, the $g_{ij}$ automatically correspond to a symplectic block structured metric, no further assumptions required (for the details see~\cite{Zen2024}).

As previously stressed~\cite{Zen2024}, BCQG comprises two foliation levels. The first level may be understood by assuming a $D$-dimensional Euclidean manifold ${\cal M}$ associated to a given analytically continuation to the complex plane metric $g^{[ac]}(x)$,  carrying coordinates $x_{\alpha}$. Following canonical steps, a preferred time-direction may be set up by defining a time function $\tau(x)$ and assign a specific time $\tau$ to each spacetime coordinate $x(\tau)$. Then we may decompose the manifold ${\cal M}$ into spatial slices $\Sigma_{\tau_i} \equiv \{x: \tau(x)= \tau_i\}$ encompassing all points $x(\tau)$ with the same ‘time-coordinate’. 
At this point, caution should be taken. This is because there is no time variable in the WDW equation in its original formulation, an intriguing feature for an equation with the ambition of describing the dynamics of quantum spacetime. Regardless of the long debates raised by the WDW equation about the nature of time, it is important to emphasize that $\tau$ in this conception should be highlighted as a non-dynamical parameter.
The gradient of the time function can be used to define a vector normal to the spatial slices, with the lapse function ensuring normalization with respect to the metric, and the lapse defined in terms of the time
coordinates $\tau$ and $\tau_{\Sigma}$ related to $\Sigma_{\tau_i}$. In the projectable Ho\v{r}ava-Lifshitz gravity, $N$ is restricted to a Euclidean time function only (see~\cite{ADM,Platania2018}). In our proposal, this restriction is overcome. In short, similarly to what occurs in general relativity based on the ADM foliation tecnique~\cite{ADM}, foliation in BCQG consists of slicing up the spacetime into three-dimensional spacelike hypersurfaces of constant time, relative to the global time function. An coordinate observer at $\Sigma_{\tau_i}$ remains his spatial coordinates constant after a time ‘lapse’ at $\Sigma_{\tau_i + d \tau_i}$. From the point of view of a Eulerian observer, that follows a normal vector orthogonal to $\Sigma_{\tau_i}$, the observer's spacelike position is shifted by a spacelike function that measures the displacement between Eulerian and coordinate observers after a time lapse, which quantifies the time interval between distinct hypersurfaces and time-dilation effects related to both observers. The second foliation level of BCQG is characterized by an analytically continued {\it Riemannian foliation} which corresponds to the reciprocal of a complex multi-valued function, the natural complex logarithm function $\ln^{-1}[\beta(t)]$, a helix-like superposition of cut-planes, which correlates Riemann sheets, with an upper edge cut in the n-th plane joined with a lower edge of cut in the (n + 1)-th plane. The BCQG universe’s scale factor $\ln^{-1}[\beta(t)]$ maps an infinite number of Riemann sheets onto horizontal strips, which represent in the branch-cut cosmology the evolution of the time-parameter dependence of horizon sizes. The patch sizes in turn maps progressively the various branches of the $\ln^{-1}[\beta(t)]$ function which are glued along the copies of each upper-half plane with their copies on the corresponding lower-half planes. In the branch-cut cosmology, the cosmic singularity is replaced by a family of Riemann sheets in which the scale factor shrinks to a finite critical size, - the range of $\ln^{-1}[\beta(t)]$, associated to the cuts in the branches, shaped by the $\beta(t)$ function —, well above the Planck length. In the contraction phase, as the patch size decreases with a linear dependence on $\ln^{-1}[\beta(t)]$, light travels through geodesics on each Riemann sheet, circumventing continuously the branch-cut, and although the horizon size scale with $\ln^{-1}[\beta(t)]^{\epsilon}$, the length of the path to be traveled by light compensates for the scaling difference between the patch and horizon sizes. Here, $\epsilon(t)$ represents the dimensionless thermodynamical connection between the energy density E and the pressure P of a perfect fluid thus enabling the fully description of the equation of state (EoS) of the system. Under these conditions, causality between the horizon size and the patch size may be achieved through the accumulation of branches in the transition region between the present state of the universe and the past events.
Conventionally, the theory deals exclusively with finite-dimensional real symplectic spaces. The BCQG in turn extends the ontological domain of general relativity to the complex plane. In BCQG, the presence of a branch-cut and a branch point define the domain of the scale factor $\ln^{-1}[\beta(t)]$ of the analytically continuous universe for the complex sector, overcoming the presence of a cosmic singularity. The BCQG offers additionally a theoretical alternative to inflation models~\cite{Guth1981,Guth2004}, based on the mathematical augmentation technique and notions of closure and existential completeness~\cite{Manders}, which have proved highly useful in both quantum mechanics~\cite{Dirac1937,Aharonov,Wu} and pseudo-complex general relativity (pc-GR)~\cite{Greiner}, with
 direct physical and cosmological manifestations. For the sake of completeness, we present in the following some elements of the BCQG formalism~\cite{Zen2024,Bodmann2023a,Bodmann2023b}.

The complete line element of the BCQG quantum gravity, resulting from the complexification of the FLRW metric~\cite{Friedman1922,Lemaitre1927,Robertson1935,Walker1937}, in line with the ADM foliation~\cite{ADM}, may be expressed as~\cite{Zen2020,Zen2021a,Zen2021b}
\begin{equation}
ds_{[\rm{ac}]}^2  \! = \!   - N^2(t) c^2dt^2  + \bigl( \ln^{-1}[\beta(t) ] \bigr)^2  \Biggl[ \!
\frac{dr^2}{\bigl(1 - kr^2(t) \! \bigr)}
    + r^2(t) \Bigl(d \theta^2 + \sin^2 \theta d\phi^2 \! \Bigr) \! \Biggr]. \label{FLRWac2}
\end{equation}
In this expression, $[\rm{ac}]$ denotes analytical continuation to the complex plane, where $r$ and $t$ represent real and complex space-time parameters, respectively, and $k$ denotes the spatial curvature of the multiverse, corresponding to negative curvature ($k = -1$), flat ($ k = 0$), or positively curved spatial hypersurfaces ($k = 1$). 
$\ln^{-1}[\beta(t)]$ represents the foliated scale factor, and
$N(t)$ denotes the lapse function. 
The gauge
invariance of the action in general relativity yields a Hamiltonian constraint that
requires a gauge-fixing condition on the lapse function (see~\cite{Feinberg}). We extend the gauge fixing constraints further to the algebraic structure of the BCQG action.

\section{Commutative BCQG}\label{CBCQG}

The starting point of this study is the  commutative BCQG action which depends on the branching scalar curvature of the universe, ${\cal R}$, and on its covariant derivatives, $\nabla^2{\cal R}, \nabla_i {\cal R}_{jk},\nabla^i {\cal R}^{jk}$, in different orders~\cite{Zen2024,Bodmann2023a,Bodmann2023b}, and whose development was based on the formulation of Ho\v{r}ava-Lifshitz~\cite{Horava}:
\begin{eqnarray}
 {\cal S}_{HL} &  =  & \int d^3x dt {\cal L} 
 =  \frac{M_P^2}{2} \int  d^3x \, dt \, N  \sqrt{g} \times \Bigl(K_{ij}K^{ij}  -  \lambda K^2   - g_0 M^2_P  
  -  g_1 {\cal R} 
  \nonumber \\ &&  -  g_ 2 M^{-2}_P {\cal R}^2  -  g_3 M_P^{-2}  {\cal R}_{ij} {\cal R}^{ij}
 -    g_4 M^{-4}_P {\cal R}^3  -  g_5 M_P^{-4}  {\cal R}({\cal R}^i_j {\cal R}^j_i) \nonumber \\ && -  g_6 M_P^{-4} {\cal R}^i_j {\cal R}^j_k {\cal R}^k_i 
-    g_7 M^{-4}_P {\cal R} \nabla^2  {\cal R} -   g_8 M^{-4}_P \nabla_i {\cal R}_{jk} 
  \nabla^i R^{jk}  \Bigr)\, .
 \label{HL} 
\end{eqnarray}
In this expression,
$g_i$ represent running coupling constants, $M_P$ denotes the Planck mass, 
and by imposing a maximum symmetric surface foliation~\cite{Zen2024},
the branching Ricci components of the three dimensional metrics 
 are given as:
\begin{equation}
    {\cal R}_{ij} = \frac{2}{\sigma^2 u^2(t)} g_{ij}\, , \quad  \mbox{and} \quad {\cal R} = \frac{6}{\sigma^2 u^2(t)} \, ,
\end{equation}
where the variable change $u(t) \equiv \ln^{-1}[\beta(t)]$, with $du \equiv d\ln^{-1}[\beta(t)]$, was introduced. 
Moreover, the trace of the extrinsic curvature tensor
$K = K^{ij} g_{ij}$, represented by $K_{ij}$~\cite{Zen2024,Bodmann2023a,Bodmann2023b,Hess}, is given as:
  \begin{equation}
K = K^{ij} g_{ij} =  - \frac{3}{2\sigma N u(t)}  \frac{du(t)}{dt}.
\end{equation}
Exploring the ADM-type formalism domain, for the metric given by equation (\ref{FLRWac2}), adopting the $N_i =0$ gauge, the three-metric $h_{ij} \to g_{ij}$ corresponds to a projectable analytically continued quantum gravity formulation: 
\begin{equation}
  h_{ij} = \ln^{-1}[\beta(t)]^2 \, {\rm diag} \Bigl(\frac{1}{1 - kr^2(t)},
    r^2(t), r^2(t)\sin^2 \theta \Bigr)  .
\end{equation}
Applying standard canonical quantization procedures to the BCQG action, the 
 canonical momentum reads
 \begin{equation}
     \pi_u = \frac{\partial S}{\partial \dot{u}} = - \frac{2 u(t) \dot{u}(t)}{N(t)} ,
 \end{equation}
and the resulting Hamiltonian density 
\begin{equation}
 {\cal H}    =   \pi_u \dot{u} - {\cal L} ,
 \end{equation}
 becomes
\begin{equation}
   {\cal H}  =  \frac{1}{2} \frac{N}{u} \Bigl[- p^2_{u} + g_r - g_m u - g_k u^2 - g_q u^3 + 
 g_{\Lambda} u^4
 + \frac{g_s}{u^2}  \Bigr] .
 \label{H*}
\end{equation} 
In this expression, the running coupling constants depict the matter-energy contributions  
of radiation ($g_r$), baryon matter ($g_m$), curvature ($g_k$), quintessence ($g_q$), cosmological constant dark energy ($g_{\Lambda})$, and stiff matter ($g_s$) (for the details see~\cite{Bertolami2011,Maeda}  and references therein).
 
\subsection{Commutative BCQG: two-fields approach}

Following similar steps to a previous contribution~\cite{Zen2024}, we introduce an extended dual field commutative manifold, composed by a mini-superspace of variables ($u(t),v(t)$), taking as a quantum counterpart-reference to the BCQG scale factor $\eta(\tau)$ the perfect fluid conception of Hermann Weyl~\cite{Weyl} denoted by $v(t)$, characterized by a dimensionless quantity $\alpha$, whose canonically conjugated momentum is defined as $p_v$. The corresponding Hamiltonian contribution is 
given as:
\begin{equation}
    {\cal H}_v \equiv \frac{1}{2} N \frac{p_v}{u(t)^{3 \alpha}}, \quad \mbox{with}
    \quad p_v = - \frac{ v(t) \dot{v}(t)}{N(t)}. \label{SHL}
\end{equation}
The variables $u(t)$ and $v(t)$ span quantum reference variable spaces.
After simplifying the notation and combining (\ref{H*}) and (\ref{SHL}), the following
super-Hamiltonian density results:
\begin{equation}
 {\cal H}   = \frac{1}{2} \frac{N}{u} \Bigl[- p^2_{u}  +    g_r  -  g_m u  -  g_k u^2  -  g_q u^3  
 +   
 g_{\Lambda} u^4   + \frac{g_s}{u^2}
    +   \frac{1}{u^{3\alpha - 1}} p_v \Bigr] \, .\label{SHuv} 
\end{equation} 
The fields $u$ and $v$ obey the following commutative Poisson algebra
\begin{eqnarray}
    \{u,v \}  =  \{ p_u, p_v \} = \{u,p_v\} = \{v,p_u\} = 0; \nonumber \\   \{u.p_u\}
    =  \{v,p_v\} = 1 \, .
\end{eqnarray}

\subsection{Commutative BCQG: three-fields extension}

In order to make contact with an inflaton-inspired proposal, we consider, in what follows, a mini-superspace triad of variables ($u(t), v(t), \phi(t)$). In this proposal, $\phi(t)$ represents a hypothetical scalar inflaton-type field, analytically continued to the complex plane, which was originally conjectured, in the real domain, by Alan Guth~\cite{Guth1981,Guth2004}, to drive cosmic inflation in the very early universe. 
According to Guth~\cite{Guth1981,Guth2004}, the main property of the physical laws that makes inflation possible is the existence of states with negative pressure, whose effects produce a repulsive form of gravity identified in the standard Friedmann equations. Based on the Robertson-Walker metric, these effects are based on an energy-momentum tensor as a function of a scalar field $\phi(x)$, the inflaton, which leads to energy and pressure densities of the universe expressed in terms of a potential, $V(\phi)$, which describes, in its original version, chaotic inflation. In the present work we also consider a simulation involving non-chaotic inflation.

\begin{figure*}[htbp]
   \centering
\includegraphics[width=0.46\textwidth]{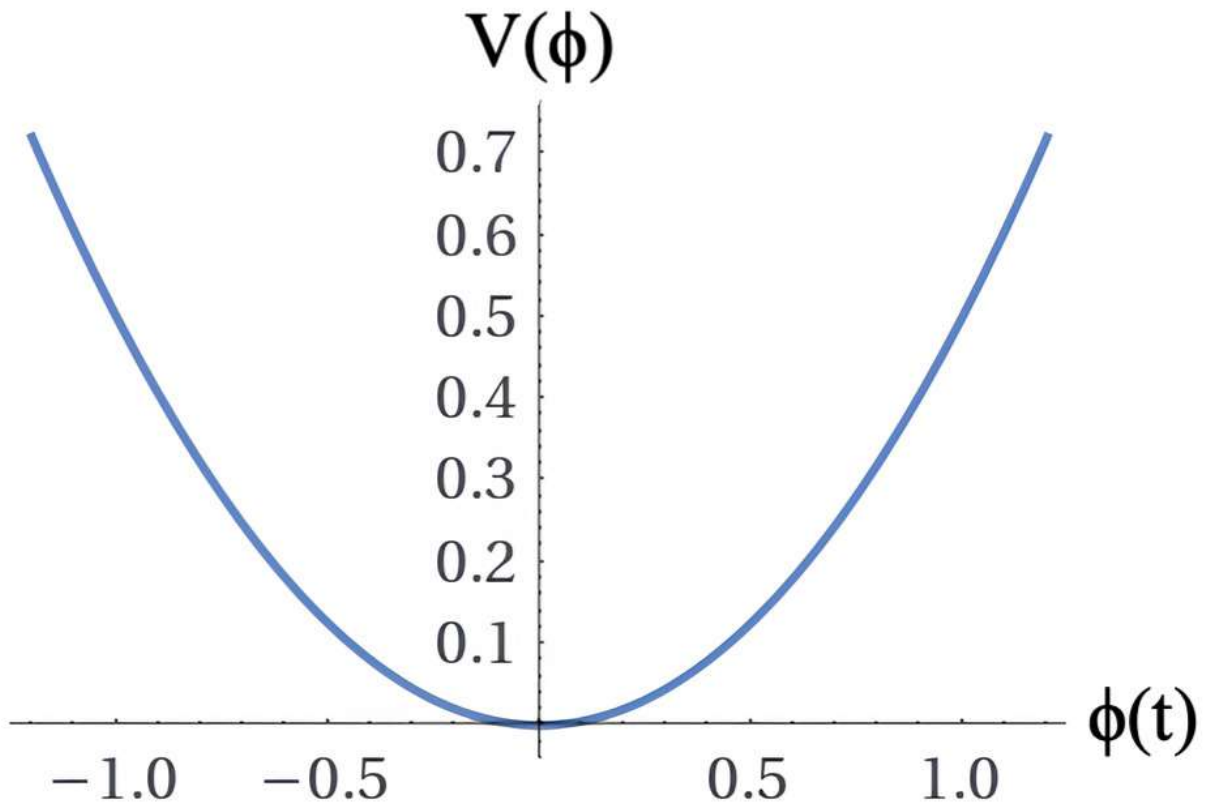}
\includegraphics[width=0.46\textwidth]{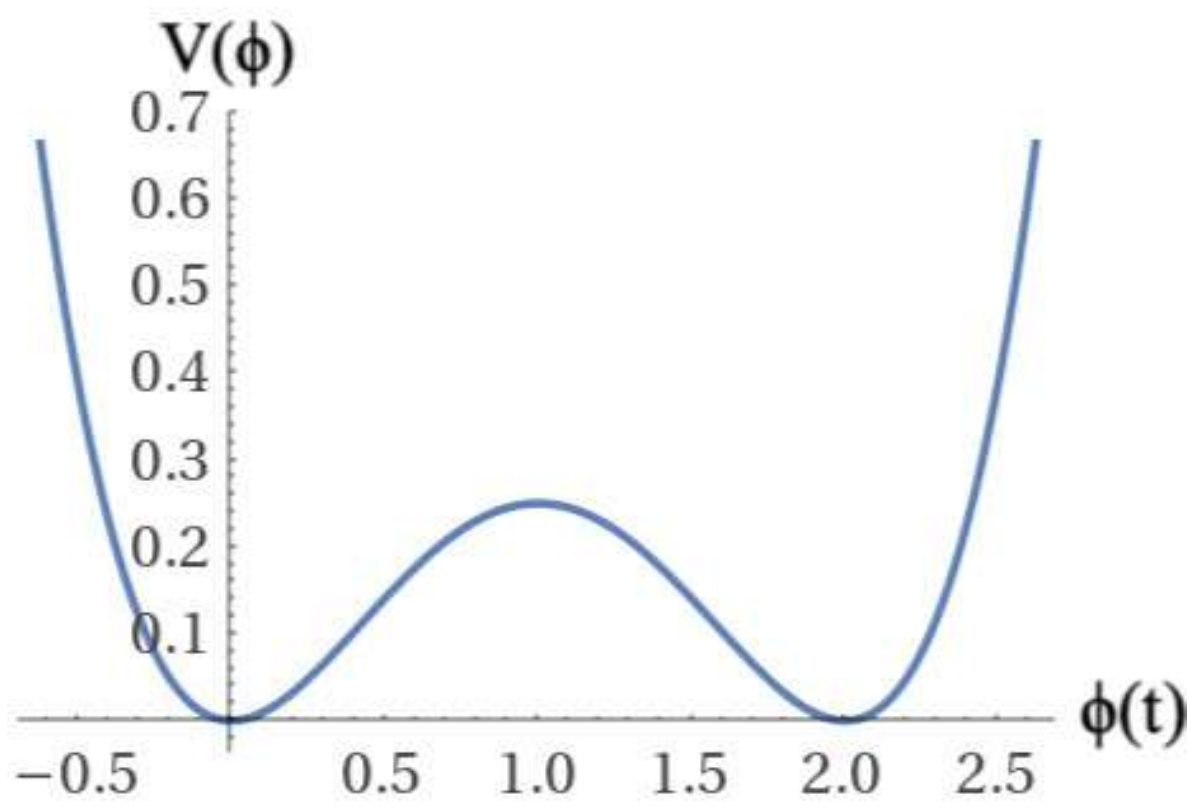}\\
    \caption{On the left, the generic form of the potential for chaotic inflation~\cite{Guth1981} indicates a relatively smooth and gradual rise. On the right, the Fubini potential~\cite{Fubini} for non-chaotic inflation shows a more intricate shape, suggesting that the scalar field dynamics in non-chaotic scenarios may involve more entangled and nonlinear behavior. }
  \label{inflaton}
\end{figure*}

We adopt for the action of the scalar field $\phi$ the following expression~\cite{Kiritsis,Tavakoli}
\begin{equation}
    S_{\phi} = \int_{{\cal M}} d^3 x \, dt \, N \sqrt{g} \Biggl(\frac{1}{2} \Bigl(\frac{1}{N^2}\Bigr) F(\phi) \dot{\phi}^2 - V(\phi) \Biggr) , \label{Sphi}
\end{equation}
where $V(\phi)$ denotes the inflation potential, defined in terms of the inflaton-type field and where $F(\phi)$ represents a coupling function.
In the following, by inserting into the BCQG action (\ref{SHL}) the scalar field contribution (\ref{Sphi}), by applying conventional quantum field theory techniques,
the corresponding contribution associated to the $\phi(t)$ field is given as
\begin{equation}
    {\cal H}_{\phi} \equiv \frac{1}{2} N \Biggl( \frac{p^2_{\phi}}{u^{3 \omega}(t)F(\phi)} + \frac{2}{u} V(\phi) \Biggr), \quad \mbox{with}
    \quad p_{\phi} = - \frac{ \phi(t) \dot{\phi}(t)}{N(t)}, \label{SHLphi}
\end{equation}
the Hamiltonian associated with the extended mini-superspace of variables may be obtained. The resulting commutative Hamiltonian density then becomes
\begin{eqnarray}
    {\cal H} & = & \frac{1}{2} \frac{N}{u} \Biggl[- p^2_u + g_ r - g_m u -g_k u^2 - g_q u^3 + g_{\Lambda} u^4  + \frac{g_s}{u^2} + \frac{1}{u^{3 \alpha - 1}} p_v \nonumber \\
    && 
    + \frac{1}{u^{3\omega - 1}F(\phi)}p^2_{\phi} + 2 V(\phi) \label{Hcomplete}
    \Biggr].
\end{eqnarray}
In what follows, we model chaotic inflation by adopting the potential~\cite{Guth1981} (see Figure \ref{inflaton})
\begin{equation}
    V(\phi) = \frac{1}{2} g^2_{\phi} \phi^2,
\end{equation}
and for the non-chaotic inflation, the Fubini potential~\cite{Fubini} (see Figure \ref{inflaton})
\begin{equation}
    V(\phi) = \frac{g^2_{\phi}}{4} (\phi - \phi_c)^4 - \frac{g^2_{\phi}}{2} (\phi - \phi_c)^2 +  \frac{g^2_{\phi}}{4}.
\end{equation}

In the following, we build a non-commutative three fields approach.
\section{Extended BCQG symplectic non-commutative algebra} 
\def\ii{\'\i}

In order to extend the BCQG symplectic non-commutative algebra, 
we embrace a similar analytical strategy adopted in the elaboration of the non-commutative two-fields BCQG. We base our strategy firstly on the construction of an extended three-field commutative BCQG manifold, composed of an extended mini-super-space of variables ($u(t), v(t), \phi(t)$). Then, this extended commutative manifold is related to a non-commutative structure, based on an extended Faddeev-Jackiw symplectic-type phase space transformation applied to the mini-super-space of variables $u(t), v(t), \phi(t)$. The insertion of the field $\phi(t)$ in the non-commutative symplectic formulation substantiates this way the extension of the BCQG scope in order to make contact with standard inflation approaches.

\subsection{BCQG as a gauge theory} 

From a conceptual point of view, it is important to emphasize that BCQG formulation essentially constitutes a quantum gauge field theory. Characterized by a group of transformations of the field variables, known as gauge transformations, a quantum gauge theory postulates the Lagrangian and the dynamics of a quantum system are invariant under local transformations, in accordance with  Lie algebra which encompasses fields and group generators. This condition, called gauge invariance principle, embraces the theory with a set of symmetries that fundamentally govern its dynamical equations. 

Among the theoretical approaches commonly adopted to systematically identify symmetries and conserved quantities in a field theory, the Dirac formalism~\cite{Dirac1964} and the Faddeev–Jackiw method~\cite{Faddeev1967,Faddeev1988} stand out. In order to accomplish conventional and elementary quantization of a dynamical system, the Dirac method for a given Lagrangian demands (i) to eliminate velocities in favor of momenta by adopting a Legendre transform that determines the corresponding Hamiltonian, (ii) to postulate canonical brackets among coordinates and momenta and finally (iii) determine the dynamics by imposing commutation rules encompassing the Hamiltonian. However, this procedure may fail when dealing with any singular Lagrangian and constrained dynamics~\cite{Jackiw}. The alternative approach of Faddeev–Jackiw~\cite{Faddeev1967,Faddeev1988}, based on Darboux's theorem, provides a symplectic and straightforward effective instrument to deal with gauge constrained systems and makes it possible to obtain the desired results — formulas for brackets and for the Hamiltonian — while avoiding canonical restrictions of the Dirac procedure. 

In what follows, we will adopt an unorthodox perspective by proposing, as previously carried out~\cite{Zen2024}, a reverse transformation from a set of non-commutative coordinates to commutative variables based on the Faddeev-Jackiw symplectic approach~\cite{Faddeev1967,Faddeev1988}, extended to the triad $u(t)$, $v(t)$ and $\phi(t)$ assumed as dual an complementary quantum fields. 
In summary, following a procedure adopted previously~\cite{Zen2024}, we adopt a reverse logic of the conventional Faddeev-Jackiw formalism, presupposing a character change for the variables $u(t)$, $v(t)$ and $\phi(t)$, assuming as stressed before that they obey the non-commutative algebraic structure described by a non-zero Poisson bracket formalism.

We assume the extended domain of commutating ($\tilde{x}_i$) and non-commuting ($x_i$)
coordinates are represented in general as
\begin{equation}
({\tilde x}_i)  =  ({\tilde u}, \tilde{p}_u, {\tilde v}, {\tilde p}_v, {\tilde \phi}, {\tilde p}_\phi ),
\quad
(x_i)  =  (u,p_u , v, p_v, \phi , p_\phi).
\label{eq-1}
\end{equation}

The commuting (on the left) and non-commuting (on the right) coordinates satisfy the following
Poisson-brackets
\begin{equation}
\mbox{commuting} \to  \left\{ {\tilde x}_i , {\tilde x}_j \right\}  =  {\tilde g}_{ij},
\quad
\mbox{non-commuting} \to 
\left\{ x_i , x_j \right\}  =  g_{ij};
\label{eq-2}
\end{equation}
the bracket on the left satisfies the commutative metric while the bracket on the right satisfies the symplectic metric, according to the following symmetry properties:
\begin{equation}
\mbox{commuting} \to {\tilde g}_{ji}  =  -{\tilde g}_{ij},
\quad
\mbox{non-commuting} \to g_{ji}  =  -g_{ij}.
\label{eq-3}
\end{equation}
The algebraic matrix structure of the commuting brackets on the left
may be cast as 
\begin{equation}
\left({\tilde g}\right)  = 
\left(
\begin{array}{cccccc}
0 & 1 & 0 & 0 & 0 & 0 \\
-1 & 0 & 0 & 0 & 0 & 0 \\
0 & 0 & 0 & 1 & 0 & 0 \\
0 & 0 & -1 & 0 & 0 & 0 \\
0 & 0 & 0 & 0 & 0 & 1 \\
0 & 0 & 0 & 0 & -1 & 0  
\end{array}
\right)
\end{equation}
Using the left expression of equation (\ref{eq-2}), the algebraic matrix-structure above leads to the following non-zero commutative Poisson-brackets:
\begin{eqnarray}
\{ {\tilde u}, {\tilde p}_u \} & = & 1; \quad
\{ {\tilde u}, {\tilde p}_v \} =  0; \quad
\{ {\tilde u}, {\tilde p}_\phi \} = 0; \quad
\left\{ {\tilde v}, {\tilde p}_u \right\} = 0; 
\nonumber \\
\{ {\tilde v}, {\tilde p}_v \} & = & 1; \quad
\{ {\tilde v}, {\tilde p}_\phi \} = 0; \quad
\{ {\tilde \phi}, {\tilde p}_u \} =  0; \quad
\{ {\tilde \phi}, {\tilde p}_v \} = 0; 
\nonumber \\
\{ {\tilde \phi}, {\tilde p}_{\phi} \} & = & 1; \quad
\{ {\tilde p}_u, {\tilde p}_v \} =0; \quad
\{ {\tilde p}_u, {\tilde p}_{\phi} \}= 0; \quad
\{ {\tilde p}_v, {\tilde p}_{\phi} \} =0.
\label{eq-5a}
\end{eqnarray}
Symplectic geometry, in turn,  whose origins date back to the Hamiltonian formulation of classical mechanics, is a branch of differential geometry and differential topology that studies symplectic manifolds, or more specifically, differentiable manifolds equipped with a closed non-degenerate 2-form.
Using the symmetry properties synthesized in the expression on the right of equation (\ref{eq-3}),
the matrix representation of a symplectic algebra may be cast as
\begin{equation}
\left(g\right)  = 
\left(
\begin{array}{cccccc}
0 & 1 & 0 & \gamma & 0 & \beta \\
-1 & 0 & -\chi & \alpha_1 & -\delta & \alpha_2 \\
0 & \chi & 0 & 1 & 0 & \zeta \\
-\gamma & -\alpha_1 & -1 & 0 & -\varsigma & \alpha_3 \\
0 & \delta & 0 & \varsigma & 0 & 1 \\
-\beta & - \alpha_2 & -\zeta & -\alpha_3 & -1 & 0   
\end{array}
\right).
\label{eq-4}
\end{equation}
For the non-commuting Poisson brackets, using equation (\ref{eq-2}), we obtain 
\begin{eqnarray}
\{ u, p_u \} & = & 1; 
\quad
\{ u, p_v \}  =  \gamma; 
\quad
\{ u, p_\phi \} =  \beta; \quad
\{ v, p_u \}  =  \chi; 
\nonumber \\
\{ v, p_v \} & = & 1; \quad
\{ v, p_\phi \} =  \zeta; \quad
\{ \phi, p_u \} = \delta; \quad
\{ \phi, p_v \}  =  \varsigma; 
\nonumber \\
\{ \phi, p_\phi \} & = & 1; \quad
\{ p_u, p_v \} =  \alpha_1; \quad
\{ p_u, p_\phi \} =  \alpha_2; \quad
\{ p_v, p_\phi \} =  \alpha_3.
\label{eq-5b}
\end{eqnarray}

In order to perform the three-field extension within the framework of non-commutative Riemannian foliated branch-cut quantum gravity (BCQG), based on an improved deformation of the conventional Poisson algebra, we connect the commutative algebraic structure with its non-commutative counterpart, by means of a symplectic topological manifold linear transformation:
\begin{equation}
x_i  =  \sum_j M_{ij} {\tilde x}_j. 
\label{eq-6}
\end{equation}
Following the steps outlined on equations (\ref{eq-7}), (\ref{eq-8}), and (\ref{eq-8-0}), we arrive at the non-commutative symplectic Poisson-type brackets synthesized in equation
(\ref{eq-8b}).
Applying the standard gauge fixing condition which states that the structure of the original equations remains invariant under coordinate transformations, so as to produce a result that is physically meaningful, we adopt in what follows the following settings for determining the matrix elements $M_{ij}$:
\begin{enumerate}
    \item[(i)] The resulting expressions for the transformations given in equation (\ref{eq-6}) should recover the two-fields non-commutative algebra developed in the previous paper~\cite{Zen2024} in case 
the set of commutative and non-commutative fields obey the condition $\phi = \tilde{\phi} = p_{\phi} = \tilde{p}_{\phi} = 0$:
    \begin{equation}
        p_u = \tilde{p}_u - \chi \tilde{p}_v; \quad p_v = - \gamma \tilde{p}_u + \alpha \tilde{u} + p_v - \alpha \tilde{v};  
    \end{equation}
\item[(ii)] The resulting transformation expressions (\ref{eq-6}) should keep the original functional structure of the super-Hamiltonian, as a kind of ‘canonical structural gauge’,  maintaining preserved the identities of the components composition, --- coordinate power series disassociated from conjugate momenta contributions ---, in short, avoiding mixing original algebraic elements through the adopted symplectic transformations, preserving the physical content logic related to the Hamiltonian formulation.
\item[(iii)] First-order canonical momentum transformations are not subject to such constraint, since first-order linear momentum dependent terms do not alter the structure of the original Hamilton equations. This is because for transformations which contains linear momentum algebraic components combined with power series involving the fields $u$, $v$, and $\phi$, the components of these power series contributions are `naturally' incorporated into the different dynamical potentials, preserving the formal structure of the corresponding Hamilton's equation.
\item[(iv)] According to ~\cite{Silva}, symplectic reduction is at the heart of many symplectic arguments. In line with the Marsden-Weinstein-Meyer Theorem,
which underpins the statement that whenever there is a symmetry group of dimension $k$ acting on a mechanical system, the number of degrees of freedom for the position and momenta of the particles can be reduced by $2k$, symplectic reduction provides extremely symmetric symplectic manifolds.
Moreover, from the previous two-field non-commutative algebra approach~\cite{Zen2024} we learn, assuming a kind of `symplectic gauge-reduction' proposal, that the symmetries of a Hamiltonian dynamical system enables the reduction of the number of degrees of freedom of a given system, generating a lower ordering dimensional symplectic manifold, which in order to be so representative as possible, it demands to settle a symmetric and equalized representation; this corresponds to an additional demand, strictly followed in the adopted formulation, as we will see below.
\end{enumerate}
An algebraic transformations that obeys all the previous condition does not change the non-commutative nature of the symplectic general algebraic structure, and obeys the fundamental gauge principle which states that the Lagrangian, and therefore the dynamics of the system itself, do not change under local transformations according to smooth families of operations of the Lie group.
Applying all these conditions we arrive at the following restrictions for the $M_{ij}$ components:
\begin{eqnarray}
    M_{21} & = & M_{23} =  M_{25} = 0; \quad   M_{22} = 1 ; \quad M_{24} = -\chi; \quad   
   M_{26} = \delta; \nonumber \\ 
 M_{41} & = & - M_{43} =  M_{45} = 
 \alpha; \quad
 M_{42} = -\gamma; \quad  
   M_{44} = 1; \quad
 M_{46} = -\varsigma; 
 \nonumber \\
 M_{61} & = & M_{63} = M_{65} = 0; \quad 
   M_{62} = -\beta; \quad 
 M_{64} = \zeta; \quad  M_{66} =  1. 
\end{eqnarray}
These conditions reduce the original symplectic matrix to the following form:
\begin{eqnarray}
\centering
\left( M_{Reduced}\right)  & = &
\left(
\begin{array}{cccccc}
1 & 0 & 0 & 0 & 0 & 0 \\
0 & 1 & 0 & - \chi & 0 & \delta  \\
0 & 0 & 1 & 0 & 0 & 0 \\
\alpha & - \gamma & -\alpha & 1 & \alpha & -\varsigma \\
0 & 0 & 0 & 0 & 1 & 0 \\
0 & -\beta & 0 & \zeta & 0 & 1 \\
\end{array}
\right).
\end{eqnarray}

The topological-algebraic constraints  previously approached result in the following transformed coordinates:
\begin{equation}
    p_u = \tilde{p}_u - \chi \tilde{p}_v + \delta \tilde{p}_{\phi}; \quad p_v = -\gamma \tilde{p}_u + \alpha \tilde{u} - \alpha \tilde{v} + \alpha \tilde{\phi} + \tilde{p}_v - \varsigma \tilde{p}_{\phi}; \label{pupv}
\end{equation}
and
\begin{equation}
    p_{\phi} =- \beta \tilde{p}_u + \zeta \tilde{p}_v + \tilde{p}_ {\phi}. \label{pphi}
\end{equation}
An important aspect to be highlighted is that, although the placed impositions imply significant formal simplifications, the most distinguishing attributes of the non-commutative nature of the original equations remain consistently incorporated. Still, the merits regarding the rigor of the analytical foundations applied in the present formulation, with respect to the implications of the non-commutative structure as inducing the acceleration of the universe, remain pertinent.

\subsection{The role of time in BCQG}

When tracing the genesis of the problem of time in quantum mechanics, Jan Hilgevoord based his observations on the work of six of the founding fathers of modern quantum theory: Dirac, Heisenberg, Bohr, Schrödinger, von Neumann and Pauli, covering the period from 1925 to 1933, finding some coherence issues between the views of these authors. In this study, a mixture of notions from classical, quantum mechanics and the theory of relativity were found~\cite{Jan}. Although this topic is not the focus of the present work, it serves as an example that the problem of time in quantum mechanics, as well as in general relativity and quantum gravity is not new. And apparently we are far from finding a final definition to this issue. In this subsection we address some aspects involving the problem of time in QCG in order to contribute to its elucidation. 

Time in quantum theory corresponds to an external element that, in a certain way, acts as a controller of all movement, identified either as an absolute quantity (similarly to Newtonian physics) or in the form of proper times characterized by a classical space-time metric, applicable to local quantum systems along their world lines. In other words, the reading of time corresponds to a classical clock underlying the quantum world. (For an useful discussion see~\cite{Zeh,Rovelli2019}). 

The dynamics of quantum systems in turn consists of individually undetermined stochastic 'quantum jumps' between 'stationary' states, or more properly, energy eigenstates~\cite{Pauli}, although time corresponds to a dynamic variable external to quantum systems flowing continuously. In this sense, the flow of time in non-relativistic quantum mechanics acquires an universal and absolute character, similarly to Newtonian mechanics.  

In quantum relativistic theory, proper times assume the role of absolute time for local systems, i.e., systems that follow, albeit approximately, the world lines in spacetime. However, since quantum states are generically non-local, auxiliary time coordinates corresponding to arbitrary foliations of spacetime can then be introduced to define the dynamics of global states. This aspect of relativistic quantum mechanics is similar to the role of the t parameter adopted in BCQG.

In quantum gravity, time as an absolute quantity has no meaning. Therefore, by obeying the Wheeler-DeWitt equation, known as the timeless equation, the wave function of the universe corresponds to a functional in a configuration space consisting of spatial geometries and matter field configurations. In this ambiance, all dynamics is encoded in the matter field configuration entanglement underlying the universal configuration space, the `DeWitt-superspace'~\cite{Zeh}. 

Therefore, the evolutionary process referred to in quantum gravity in general, as well as in BCQG, does not allude to an `absolute' time as a dynamic variable, but to the evolution of the different configurations of matter fields. It is precisely for this reason that the evolutionary equations of the wave function of the universe refer to derivatives in terms of the fields $\eta$, $\xi$, and $\varphi$, and not to temporal derivatives. Furthermore, the presence of the variables $t$ in the mentioned equations refers in BCQG to parametric quantities related to spacetime foliations, making it possible to associate the term `evolution' with a process connected to the systematically  unfolding transformations of matter field configurations on the different Riemannian sheets.
\subsection{Probability interpretation of the wave function of the universe}

The evolving universe is described in quantum gravity by a wave-function that corresponds to solutions of the Wheeler-DeWitt functional differential equations and appropriate boundary conditions.
In standard quantum mechanics, the probability to find a given system characterized by a wave function $\Psi(q_i,t)$ in a configuration-space element $d\Omega_q$ at time $t$ is given by
\begin{equation}
dP = |\Psi(q_i,t)|^2 d\Omega_q .
\end{equation}
In case $\Psi(q_i,t)$ is well behaved at infinity, then the above integral, taken over the whole configuration space, is independent of time and can be normalized to one
(for an interesting discussion of this topic see~\cite{Hawking1982,Vilenkin,Vilenkin1988}).

In quantum gravity, the wave-function of the universe, $\Psi(h_{ij}(\mathbf{x}), \Phi(\mathbf{x}),\Sigma(\mathbf{x}))$, as stressed before, corresponds to a functional 
on the geometries of compact manifolds $\Sigma(\mathbf{x})$, defined in terms of a three-dimensional metrics superspace, $h_{ij}(\mathbf{x})$, and matter field configurations, $\Phi(\mathbf{x})$~\cite{Hawking1982,Vilenkin1988,Rovelli2004,Rovelli2011,Rovelli2015}. 
According to 
Hawking and Page~\cite{Page},   the 
probability $P({\cal A})$ of finding a 3-surface $\Sigma$ with metric $h _{ij}$ and matter field configuration $\Phi$ is given as:
\begin{equation}
P({\cal A}) \propto \int_{\cal A} |\Psi[h _{ij},\Phi,\Sigma]|^2 \, d{\cal V} , \label{HawkingPrescription}
\end{equation}
where $d{\cal V}$ corresponds to a volume element.
The replacement of the time evolution of the wave function of the universe by the (dynamical) evolution of the matter field configuration and, therefore, the absence of time in the conventional sense in the framework of quantum gravity is a characteristic of the classical Hamilton-Jacobi formulation of general relativity~\cite{Rovelli2015}.
In what follows, we adopt the term `probability density' to establish a connection involving the BCQC and conventional quantum mechanics.

\section{The wave function of the universe in the non-commutative three-fields formalism} 
As a result of the previously adopted gauge-transformations, from equation (\ref{Hcomplete}) combined with (\ref{pupv}) and (\ref{pphi}) we arrive at the following super-Hamiltonian:
\begin{eqnarray}
 {\cal H} 
       & = &  \frac{1}{2}\frac{N}{{u}} \Bigl[- \Bigl({p}_u  - \chi {p}_v + \delta {p}_{\phi}\Bigr)^2
       -  \frac{1}{u^{3 \alpha-1}} \Bigl(\gamma {p}_u - \alpha {u}- {p}_v  + \alpha {v} + \varsigma p_{\phi} - \alpha \phi \Bigr)   
       \nonumber \\
       && +  \frac{1}{u^{3 \omega-1}F(\phi)} \Bigl( 
      - \beta p_u + \zeta p_v + p_{\phi} \Bigr)^2 + 2 V(\phi) 
       \nonumber \\
&& + \Bigl( g_r - g_m {u} -  g_k {u}^2 - g_q {u}^3 +     
 g_{\Lambda} {u}^4 
    + \frac{g_s}{{u}^2}  \Bigr) \Bigr] \, . 
    \label{HTinverted} 
\end{eqnarray} 
In this expression, for notation simplicity, we eliminate the tilde identification of the commutative variables. 
By adopting the reverse mapping path proposal, the above equation materializes
the effects of reconfiguration of the original super-Hamiltonian through the imposition of a non-commutative algebra. The resulting equation, although dependent on the original commutative variables, highlights this reconfiguration through the imposition of a structural composition that inserts new dynamic components into the original formalism, modulated by the parameters $\sigma$, $\chi$, $\gamma$, $\alpha$.  
Unlike a conventional symplectic transformation,  this procedure makes it possible to identify, in a comprehensible way, the striking outcome of the non-commutative algebraic transformations when compared to a standard formulation. 

From expression (\ref{HTinverted}), making the quadratic terms explicit, we obtain
\begin{eqnarray}
 {\cal H} 
       & = &  - \frac{1}{2}\frac{N}{{u}} \Bigl[\Bigl( {p}^2_u +\chi^2 {p}^2_v + \delta^2 {p}^2_{\phi} - 2 \chi p_u p_v + 2 \delta p_u p_{\phi} - 2 \chi \delta p_v p_{\phi}\Bigr) \nonumber \\
&&       +  \frac{1}{u^{3 \alpha-1}} \Bigl(\gamma {p}_u - \alpha {u}- {p}_v  + \alpha {v} + \varsigma p_{\phi} - \alpha \phi \Bigr)   
       \nonumber \\
       && -  \frac{1}{u^{3 \omega-1}F(\phi)} \Bigl( 
      \beta^2 p^2_u + \zeta^2 p^2_v + p^2_{\phi} -2\beta \zeta p_u  p_v - 2 \beta p_u p_{\phi} + 2 \zeta p_v p_{\phi}\Bigr)  \nonumber \\
&&- 2 V(\phi) 
       - \Bigl( g_r - g_m {u} -  g_k {u}^2 - g_q {u}^3 +     
 g_{\Lambda} {u}^4 
    + \frac{g_s}{{u}^2}  \Bigr) \Bigr] \, . 
    \label{HTinverted+} 
\end{eqnarray} 
\begin{table}[htbp]
    \centering
    \begin{tabular}{|c|c|c|c|c|c|} \hline
        ${\cal C}_1$ &  ${\cal C}_2$ &  ${\cal C}_3$ &  ${\cal C}_4$ &  ${\cal C}_5$ &  ${\cal C}_6$ \\
        \hline
      $1 - \beta^2$  &  $\chi^2 - \zeta^2$ & $\delta^2 - 1$ & $2 \beta \zeta -2\chi$ & $2 \delta + 2 \beta$ & $- 2 \chi \delta - 2 \zeta$ \\ \hline
    \end{tabular}
    \caption{Assuming the perfect fluid radiation condition for the early universe matter content,
$\omega = 1/3$ and the naturalness condition, the corresponding expressions for the coefficients
$C_i$ (i = 1, ..., 6) are shown in the table. Making the quadratic terms explicit in equation 
(\ref{HTinverted++}) and  combining similar terms, the coefficients
$C_i$ comprise the parametric contributions.}
    \label{Ci}
\end{table}
Assuming the perfect fluid radiation condition for the early universe matter content, $\omega = 1/3$ and naturalness condition, the corresponding expressions for the coefficients ${\cal C}_i(i = 1,...,6)$ are shown in table \ref{Ci}. 
Combining similar terms, from the previous expression we obtain 
\begin{eqnarray}
 {\cal H} 
       & = & - \frac{1}{2}\frac{N}{{u}} \Bigl[\Bigl( {\cal C}_1 {p}^2_u + {\cal C}_2  {p}^2_v +  {\cal C}_3  {p}^2_{\phi} +  {\cal C}_4 p_u p_v + 
        {\cal C}_5
        p_u p_{\phi} +  {\cal C}_6 p_v p_{\phi}\Bigr) \nonumber \\
&&      + \frac{1}{u^{3 \alpha-1}} \Bigl(\gamma {p}_u - \alpha {u}- {p}_v  + \alpha {v} + \varsigma p_{\phi} - \alpha \phi \Bigr)   
      - 2 V(\phi)  \nonumber \\
&&
       - \Bigl( g_r - g_m {u} -  g_k {u}^2 - g_q {u}^3 +     
 g_{\Lambda} {u}^4 
    + \frac{g_s}{{u}^2}  \Bigr) \Bigr] \, . 
    \label{HTinverted++} 
\end{eqnarray} 

Canonical quantization procedures applied to the Hamiltonian (\ref{HTinverted++}), allow
 the variables $u(t)$, $v(t)$, and $\phi(t)$ along with their corresponding conjugate momenta $p_u$, $p_v$, and $p_{\phi}$ to be treated as operators:
 \begin{equation}
 {p}_u \to -i \frac{{\partial}}{\partial u}; \quad {p}_v \to -i \frac{{\partial}}{\partial v}; \quad \mbox{and} \quad
  {p}_{\phi} \to -i \frac{{\partial}}{\partial \phi}. \label{cv}
 \end{equation}

Using the canonical requirements given in expression (\ref{cv}) above, from equation (\ref{HTinverted++}) we obtain
the wave equation 
\begin{eqnarray}
 {\cal H} \Psi(u,v,\phi) 
       & \! = \! &   \Bigl[  \Bigl( {\cal C}_1  \frac{\partial^2}{\partial u^2}  \! +  {\cal C}_2  \frac{\partial^2}{\partial v^2} \! +  {\cal C}_3 \! \frac{\partial^2}{\partial \phi^2} \! +  {\cal C}_4  \frac{\partial}{\partial u}\frac{\partial}{\partial v}  \! +   
        {\cal C}_5 
       \frac{\partial}{\partial u}\frac{\partial}{\partial \phi}  \! +  {\cal C}_6 \! \frac{\partial}{\partial v}\frac{\partial}{\partial \phi} \Bigr) \nonumber \\
&&       +  \frac{1}{u^{3 \alpha-1}} \Bigl( i\gamma \frac{\partial}{\partial u} + \alpha {u} - i \frac{\partial}{\partial v}  - \alpha {v} + i\varsigma \frac{\partial}{\partial \phi} + \alpha \phi \Bigr)   
      + 2 V(\phi)  \nonumber \\
&&
       + \Bigl( \! g_r \! - g_m {u} -  g_k {u}^2 \! - g_q {u}^3 +     
 g_{\Lambda} {u}^4 
   \! + \frac{g_s}{{u}^2}  \! \Bigr) \! \Bigr] \! \Psi(u,v,\phi) = 0  , 
    \label{HTinverted+++} 
\end{eqnarray} 
where $\Psi(u,v,\phi)$ denotes the wave-function of the universe.  

Following the steps outlined in~\ref{B}, equation (\ref{HTinverted+++})  is reduced to the canonical form (for the details
see~\cite{Polyanin}):
\begin{eqnarray}
&& \Biggl[ \Biggl( \frac{\partial^2}{\partial \eta^2} 
  + \frac{i\gamma}{\eta^{3 \alpha-1}} \frac{\partial}{\partial \eta}  +
 g_r  - g_m \eta  -  g_k \eta^2 - g_q \eta^3   + g_{\Lambda} \eta^4 + \frac{g_s}{\eta^2}  +    \frac{\alpha}{\eta^{3\alpha-2}} \Biggr)  \nonumber
 \\
&& + \Biggl( \frac{\partial^2}{\partial \xi^2} - \frac{i}{\eta^{3\alpha-1}}\frac{\partial}{\partial \xi}  - \frac{\alpha \xi}{\eta^{3\alpha-1}}  \Biggr) \nonumber \\
&& + \Biggl( \frac{\partial^2}{\partial \varphi^2} + \frac{i\varsigma}{\eta^{3\alpha-1}}\frac{\partial}{\partial \varphi}  + \frac{\alpha \varphi}{\eta^{3\alpha-1}} 
+ 2 V(\varphi) \Biggr)\Biggr]  \Psi(\eta,\xi, \varphi)  = 0 . \label{Hsupersuper*}
\end{eqnarray}

Exploiting a relevant feature of the adopted canonical transformation (see~\ref{B}), which can be summarized in the freedom of choice of canonical variables, we opt for the following associations, $u \to \eta$, $v \to \xi$ and $\phi \to \varphi$. Evidently, the connections imposed by the relations (\ref{Q}), (\ref{Qlinear}), (\ref{Qc}) and (\ref{yi}) attribute to each variable $\eta(t)$, $\xi(t)$ and $\varphi(t)$ a new identity when compared with their original counterparts, $u(t)$, $v(t)$ and $\phi(t)$, forming a triplet of canonical complementary dual-conjugate quantum fields. Complementarity in conventional quantum mechanics is commonly interpreted in terms of duality and opposition, considering conjugate fields as dual and opposite. The formulation of quantum mechanics is shaped by complex, dual, and complementary Hilbert spaces. Quantum complementarity unifies duality and opposition in a consistent way, thus underpinning the physical world. This character of dual Hilbert spaces extends to quantum fields, when they experience sets of transformations in a non-commutative algebra, attributing to them new identities that materialize in the sharing of identities in comparison to fields originally belonging to a commutative algebraic structure. The concept of entanglement, coined in quantum mechanics to designate attributes of correlation related to conjugate variables, establishes that the quantum state of each field of a given group cannot be described independently of the state of the others, even when the particles are separated by large distances. In our view, we identify these aspects of quantum entanglement with the attributes of the present correlation, which fundamentally characterizes a non-commutative symplectic transformation of coordinates.

In the reverse Faddeev-Jackiw formalism proposed, the variables $u(t)$, $v(t)$, and $\phi(t)$ are non-commutative. After the symplectic transformation,
the new canonical variables $\xi(t)$,  $\eta(t)$, and $\phi(t)$, --- despite conforming a commutative set of variables ---, obey a set of non-commutatively structured equations, shaped by the corresponding parametric attributes of the Poisson-type non-commutative algebra. In summary, the resulting set of variables, $\eta(t)$, $\xi(t)$, and $\varphi(t)$ become canonically conjugate dual and complementary variables, which span reciprocal spaces, so the following relation between these variables holds:
\begin{equation}
\eta(t) = \frac{1}{\sqrt{2\pi}} \int_{-\infty}^{\infty} A(\xi) e^{i \eta(t) \xi(t)} d \xi; \quad \eta(t) = \frac{1}{\sqrt{2\pi}} \int_{-\infty}^{\infty} A(\varphi) e^{i \eta(t) \varphi(t)} d \varphi,
\end{equation}
and 
\begin{equation}
\xi(t) = \frac{1}{\sqrt{2\pi}} \int_{-\infty}^{\infty} A(\varphi) e^{i \xi(t) \varphi(t)} d \varphi. 
\end{equation}

The ideal fluid condition for the radiation era, assuming the $\alpha$ parameter is a real quantity,  $\alpha = 1/3$, allows the variables separation of equation (\ref{Hsupersuper*}) in the form
\begin{eqnarray}
&& \Biggl( \frac{\partial^2}{\partial \eta^2} 
  + i\gamma \frac{\partial}{\partial \eta}  + {\cal V}(\eta) 
  \Biggr) \Psi(\eta)  =  0; \label{psieta}
\\
&& \Biggl(\frac{\partial^2}{\partial \xi^2} -  i\frac{\partial}{\partial \xi}  - \frac{\xi}{3}  \Biggr) \Psi(\xi) 
 = 0 ; \label{psixi} \\
&& \Biggl(\frac{\partial^2}{\partial \varphi^2} + i\varsigma \frac{\partial}{\partial \varphi}   +  \frac{\varphi}{3}  + 2 V(\varphi) \Biggr)  \Psi(\varphi)  =  0 ,
\label{psivarphi}
\end{eqnarray}
with $\Psi(\eta,\xi,\varphi) \equiv \Psi(\eta)\Psi(\xi)\Psi(\varphi)$, $g_m^{\prime} \equiv g_m - 1/3$ and
\begin{equation}
{\cal V}(\eta) \equiv g_r  - g_m^{\prime} \eta  -  g_k \eta^2 - g_q \eta^3   + g_{\Lambda} \eta^4 + \frac{g_s}{\eta^2}.  
\end{equation}
The parameters $\gamma$ ans $\varsigma$ may be assumed as real or imaginary quantities. In the first case, equations (\ref{psieta}) and (\ref{psivarphi})
may be cast as
\begin{eqnarray}
&& \Biggl( \frac{\partial^2}{\partial \eta^2} 
  + i|\gamma| \frac{\partial}{\partial \eta}  
 + {\cal V}(\eta)   \Biggr) \Psi(\eta)  =  0; \label{psieta+}
\\
&& \Biggl(\frac{\partial^2}{\partial \varphi^2} + i|\varsigma| \frac{\partial}{\partial \varphi}   +  \frac{\varphi}{3}  + 2 V(\varphi) \Biggr)  \Psi(\varphi)  =  0 .
\label{psivarphi+}
\end{eqnarray}
In the second case, assuming
$\gamma = i|\gamma|$ and $\varsigma = i|\varsigma|$, equations (\ref{psieta}) and (\ref{psivarphi}) become:  
\begin{eqnarray}
&& \Biggl( \frac{\partial^2}{\partial \eta^2} 
  - |\gamma| \frac{\partial}{\partial \eta}  
 + {\cal V}(\eta)   \Biggr) \Psi(\eta)  =  0; \label{psieta++}
\\
&& \Biggl(\frac{\partial^2}{\partial \varphi^2} -|\varsigma| \frac{\partial}{\partial \varphi}   +  \frac{\varphi}{3}  + 2 V(\varphi) \Biggr)  \Psi(\varphi)  =  0 ,
\label{psivarphi++}
\end{eqnarray}

\subsection{Foliated branch-cutting approaches to classical and quantum gravity}

Before performing calculations and interpreting the results, it is important to make a distinction between the classical approach of the branch-cut foliate formulation and its quantum counterpart. The classical formulation was based on S. Hawking and T. Hertog's multiverse proposal~\cite{Hawking2018} of a hypothetical set of multiple universes existing in parallel and on the analytic continuation technique in complex analysis applied to the Friedmann-Lemaître-Robertson-Walker (FLRW) metric~\cite{Friedman1922,Lemaitre1927,Robertson1935,Walker1937}. As a result of this approach, a closed system of Friedmann-type field equations that sweeps a hypothetical maximally symmetric and homogeneous set of superposed multiple universes, with a cosmic scale factor analytically continued into the complex plane, was obtained. Complexification of the FLRW metric resulted in a linearly independent superposition of these field equations associated with infinitely many poles, in line with Hawking's multiverse composition of an infinite number of simultaneously occurring primordial universes arranged along a line in the complex plane with infinitesimal residues. The introduction of a regularization variable allows shifting the limits of the Friedmann-type field equations beyond the primordial singularity.
Imposing in turn that the multiple singularities of the field equations are confined to the same universe and using a Riemann integration, a branch-cut solution was obtained.
The regularization function allows the contour solution-lines to move around the branch cut, since the integration limits can be shifted without altering the continuity of the resulting functions so long as the contour-lines does not cross the complex branch-point related to the branch-cut.
The introduction of a regularization function at this stage of the formulation is not equivalent to changing the limits of the integration of Friedmann’s equations to avoid the presence of singularities, since essential or real singularities cannot be removed simply by any coordinate transformation. The technical procedure adopted admit to circumvent the singularities,~--- that would otherwise be inescapable~---, 
which in turn become branch points. To accomplish this proposal, the assumption of continuity at the local level prevails, i.e., there is some neighborhood of the branch-point, say $z_0$, close enough although not equal to $z_0$, to characterize a small region around local patches where $\ln^{-1}[\beta(t)]$ is single-valued and continuous. The cuts in the branch cut are shaped in turn by the function $\beta(t)$ which, besides the range, characterizes also the foliation regularization of $\ln^{-1}[\beta(t)]$
and domain extension. This procedure moreover allows a formal treatment consistent with the Planck scales that establishes, according to the multiverse concept, the region of confluence between quantum mechanics
and general relativity. The descriptive emphasis of the evolutionary process of the foliated quantum universe is focused on the imaginary sector associated with the scale factor $\ln^{-1}[\beta(t)]$. This is because each Riemann sheet of the multiple leaves associated with $\ln^{-1}[\beta(t)]$ comprises a new universe, thus composing an infinity of universes connected by a branch-cut. In the real sector, however, the different multiverses behave as disconnected, linearly independent in their evolutionary process. In this context, although the evolutionary perspective of branch-cut cosmology contemplates the imaginary sector of the universe's scale factor, the formal consistency of the proposal is also supported by the real sector.  

Concerning the quantum approach of the branch-cut foliate formulation, we adopt a topological quantization method applied to the BCQG scale factor. Topological quantization can be applied to any field configuration whose geometric structure allows the existence of a principal fiber bundle, --- a locally product space, although it may have a different topology globally. In the case of gravitational systems with an infinite number of degrees of freedom, a theorem proves the existence and uniqueness of such a bundle. According to the theorem, any solution of the minimally coupled Einstein equations to any gauge matter field can be represented geometrically as a principal fiber bundle with spacetime as the base space. The structure group, isomorphic to the standard fiber, follows from the invariance of the metric of the base space with respect to Lorentz and gauge transformations. The topological invariants of the corresponding principal fiber bundle lead to a discretization of the parameters entering the metric of the base space~\cite{Quevedo}. The quantization of the BCQG Lagrangian density is achieved by raising the dynamical variable $\ln^{-1}[\beta(t)] \to \eta(t)$ and the conjugate momentum $p_{\eta}$ to the category of quantum operators. 
Although the association between the classical variable and the quantum version are underlying present in this process of topological quantization, the formulation of BCQG  provides a new analytical conception. In this conception, the overcoming of the primordial singularity by means of a Riemannian structure that allows it to be bypassed is replaced by a quantum leap.
In other words, the primordial singularity that is replaced in the branch-cut formalism by a branch point is present in the form of an infinite succession of individual singularities, continuously interconnected, overcome however by means of quantum leaps of the solutions.

As highlighted in a previous article~\cite{Zen2024}, regarding the presence of a mirror Universe, the BCQG resembles cyclic and bouncing models that experience infinitely alternating periods of rapid expansion and contraction, surpassing the primordial singularity, without an end and without a beginning. However, when we examine 
 from an ontological and epistemological point of view 
the BCQG and compare its theoretical construction with the other formulations, the similarities between the two lines of investigation are quite remote. Cyclic bouncing models, for example, are proposed based on an analytical investigation of the evolution of the Universe and implemented mainly in an ad hoc way, through parameterizations, cosmic wedge diagrams and other aesthetic foundations. The foliated analytical continuation of quantum gravity, in turn, from an ontological and epistemological point of view, --- except for the complexification of the standard metric, combined with the concepts of multiverse ---, contemplates theoretical foundations and theoretical investigation procedures similar to those of general relativity. BCQG shares with general relativity the same fundamental questions, the same objects of investigation, the same claims about the nature of being and existence, that is, the same first principles of its conceptual philosophy. In this sense, the realization of a transition that overcomes the primordial singularity giving rise to a mirror universe is not the result of an ad hoc proposition, but the natural result of an evolutionary process of fundamental equations of the general relativity type, generated through ontological, methodological and epistemological theoretical procedures based on field theory. Likewise, the realization of the transition region that overcomes the primordial singularity does not require the imposition of an artificial mechanism to guarantee its existence, being the natural result of a topological restructuring of space-time.

For a review of these topics see~\cite{Zen2024,Zen2020,Zen2021a,Zen2021b,Zen2022,Zen2022a,Zen2023a}.

\subsection{Naturalness}

In the study of the evolution of the universe, we are faced with an epistemic limitation regarding the scientific realism, --- the under-determination of any scientific theory by evidence, a problem that particularly affects the frontiers of quantum gravity. Therefore, in order to overcome these limitations, it becomes essential to establish an organizing and guiding principle in order to make non-empirical models viable, enabling the realization of consistent and precise calculations, avoiding in particular the assumption of adjustment parameters. Proposed by Weinberg~\cite{Weinberg1972}, the principle of naturalness serves as a standard method for classifying and organizing the distinct theoretical terms of highly intricate field theoretical approaches, as well as providing guidance for understanding the various interaction couplings associated with the dynamic composition of matter, energy and even primordial sources of gravitational waves. This principle suggests that the underlying parameters in quantum field theory are all of the same ‘size’, in appropriate units or, more precisely, that a given quantum field theory can only describe nature at energies below a certain cutoff scale~\cite{Weinberg1972}. In this context, we adhere to the principles of naturalness, normalizing the coupling constants to unity.  
As an additional message, it is important to remember
the meaning of the term ‘fine-tuning’, which is used commonly to characterize sensitive dependence of facts or properties on the values of certain parameters. 
Fine tuning in particular is a requirement of standard inflation model concerning finely tuned initial conditions in order to explain the degree of flatness and homogeneity observed in the universe.
In this sense, the criterion of naturalness represents a kind of {\it no fine-tuning} condition in a rather different sense.

\subsection{Boundary conditions}

In the following we determine solutions for equations (\ref{psieta}), (\ref{psixi}), and (\ref{psivarphi}). The boundary conditions of the solutions are based on the Bekenstein criterion~\cite{Bekenstein1981}, which provides an upper limit for the universe’s entropy, following the proposition presented in~\cite{Zen2024,Bodmann2023a,Bodmann2023b}. Accordingly, a key factor to understand the upper bound of entropy contained within a certain finite region of space with a finite amount of energy is the Bekenstein bound, a fundamental criterion which settles the basis for the generalization of entropy and the second law of thermodynamics for non-gravitational systems. Applied to the primordial universe by considering the connected spatial region within
the particle horizon of a given observer, i.e., the locus
of the most distant points that can be observed at a specific time $t_0$ in an event, an upper bound, given by  $\frac{2 \pi R}{\hbar c}$, for  the entropy $S$ and energy $E$ of a system enclosed in a spherical region of radius $R$, was assumed:  
\begin{eqnarray}
    \frac{2 \pi R}{\hbar c} \geq S/E \quad \mbox{so} \quad S \leq S_B = \frac{2 \pi}{\hbar c} E R,  \label{S}
\end{eqnarray}
with $S_B$ denoting the upper entropy limit of the Bekenstein bound.  
The implications of the Bekenstein criterion are striking since it establishes initial physical conditions that impact the evolution, symmetries and conservation laws of elementary particles in the early universe. The fulfillment of the criterion would imply  a non-singular, isotropic and homogeneous primordial universe  with entropy, temperature, and baryon number equal to zero. 

The total entropy of a black hole, according to the Bekenstein criterion, is proportional to the number of Planck areas needed to cover the event horizon, where each area corresponds to one unit of entropy. In the non-commutative branched gravitation, we assume that the primordial singularity is equally covered by a certain number of Planck areas, whose numerical value in turn corresponds to the total primordial entropy of the universe (see also~\cite{Hawking1982,Hawking1985}. We also assume that the dimensions of this boundary region correspond to the most distant observable points, while respecting causality. To comply with this requirement, we consider an appropriate cosmic distance, denoted as $d(t)$, between a pair of objects at any instant of time $t$, and the corresponding distance $d(t_0)$ at a reference time $t_0$. We then establish this relation as $d(t) = |\eta(t)|d(t_0)$. This means that the relation between the two distances is modulated by the scale factor $\eta(t)$ of the BCQG universe. This implies that for $t = t_0$ we have $|\eta(t_0)| = 1$. Regarding the wave-function of the universe, adopting a conventional probabilistic point of view, this condition implies, 
at $\eta(t_0)\equiv \eta_0 = \pm 1, 0$, $|\Psi(\eta_0)|^2 = 1, 0$, assuming a normalized wave-function. Thus, the following boundary conditions can be considered: 
on the mirror universe, $\Psi(-1) = -1, \Psi'(-1) = 0$ and $\Psi(-1) = 0, \Psi'(-1) = -1$ and on the present universe, $\Psi(1) = 1, \Psi'(1) = 0$ and $\Psi(1) = 0, \Psi'(1) = 1$. 
\subsection{Solutions for $\Psi(\eta)$, $\Psi(\xi)$, and $\Psi(\varphi)$}
\subsubsection{$\Psi(\eta)$ solutions}

By mean of the successive approximation method, we found the following algebraic solution for equation (\ref{psieta+})
\begin{eqnarray}
    \Psi(\eta)  & =  &   a_1 e^{-\frac{i|\gamma|}{2}(\eta - log(\eta)/i|\gamma|)}J_{\frac{i\sqrt{3}}{2}}\Bigl(\frac{|\gamma| x}{2}\Bigr)  +  a_2 e^{-\frac{i|\gamma|}{2}(\eta - log(\eta)/i|\gamma|)}Y_{\frac{i\sqrt{3}}{2}}\Bigl(\frac{|\gamma| x}{2}\Bigr), \nonumber \\
&& + b_1 + ib_2 |\gamma| \sum_{m=1}^5 \frac{\eta^m}{m!} + 
       \frac{1}{i|\gamma|}  \sum_{n=6}^{12} b_n \eta^n  + {\cal O}\bigl(\eta^{13}\bigr), \label{psieta*} 
\end{eqnarray}
with the conditions $a_1, a_2 \to 1$ and $b_1$, $b_2$, $b_n \to 0$ as $\eta \to 0$, and $a_1, a_2 \to 0$ and $b_1$ and $b_2 \to 1$ as $\eta \to \infty$.
For equation
(\ref{psieta++}), we have obtained the following solution through the successive approximation method:
\begin{eqnarray}
    \Psi(\eta)  & =  &   a_1 e^{-\frac{|\gamma|}{2}(\eta - log(\eta)/|\gamma|)}J_{\frac{i\sqrt{3}}{2}}\Bigl(-\frac{i|\gamma| x}{2}\Bigr)  +  a_2 e^{-\frac{|\gamma|}{2}(\eta - log(\eta)/|\gamma|)}Y_{\frac{i\sqrt{3}}{2}}\Bigl(-\frac{i|\gamma| x}{2}\Bigr), \nonumber \\
&& + b_1 + b_2 |\gamma| \sum_{m=1}^5 \frac{\eta^m}{m!} + 
       \frac{1}{|\gamma|}  \sum_{n=6}^{12} b_n \eta^n  + {\cal O}\bigl(\eta^{13}\bigr), \label{psieta**} 
\end{eqnarray}
with similar conditions as the previous ones for the parameters $a_1, a_2$, $b_1$, $b_2$, and $b_n$.

\begin{figure*}[htbp]
   \centering
\includegraphics[width=0.493\textwidth]{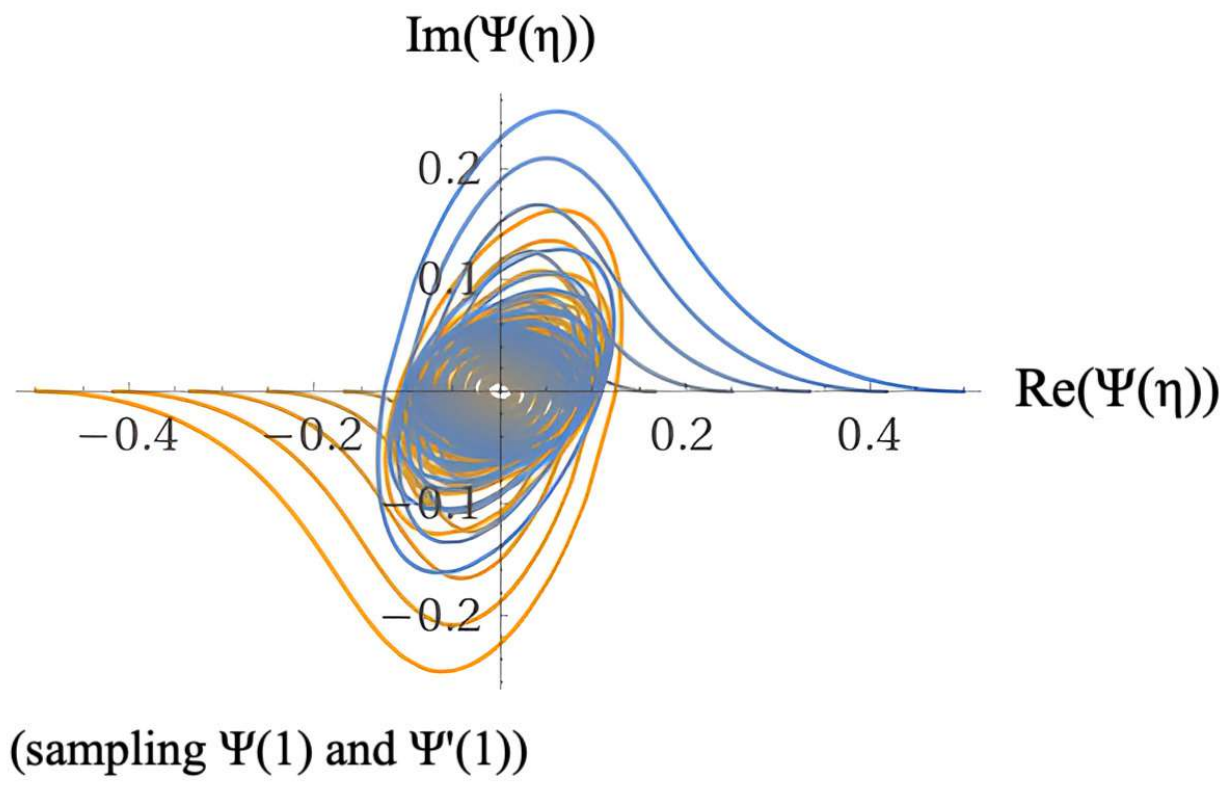}
\includegraphics[width=0.35\textwidth]{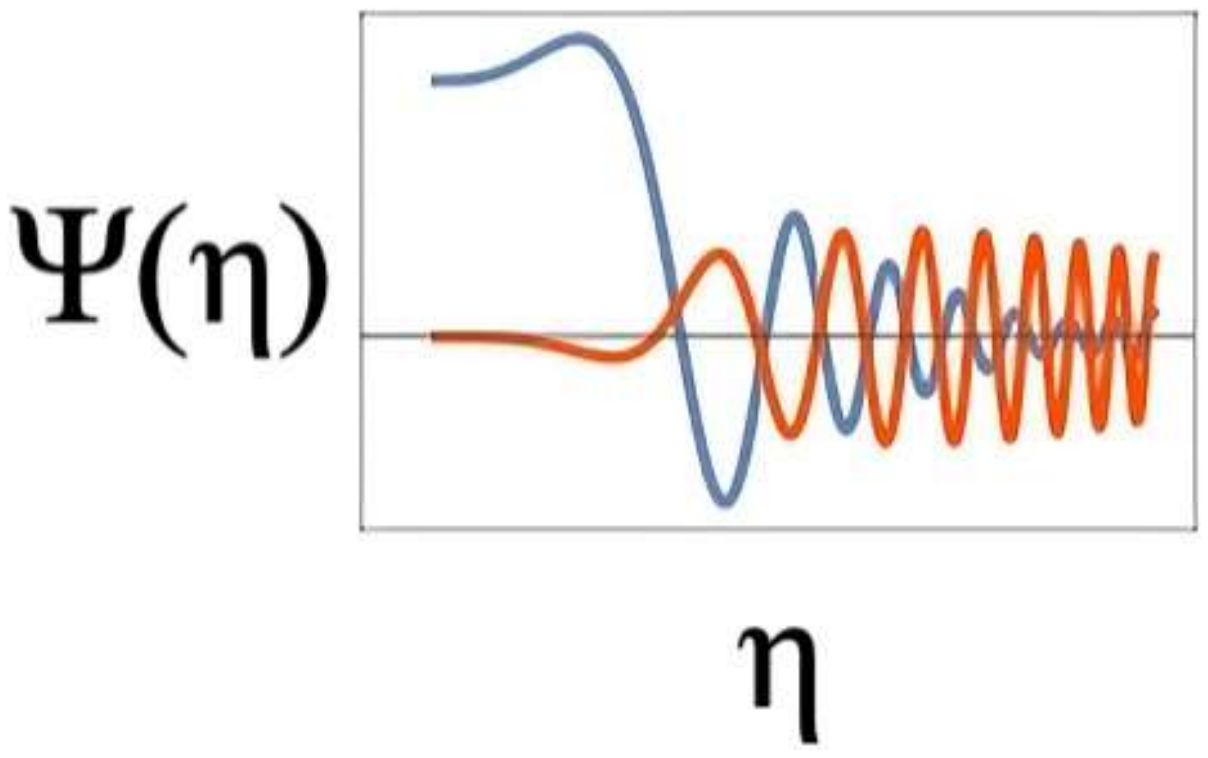}
\includegraphics[width=0.35\textwidth]{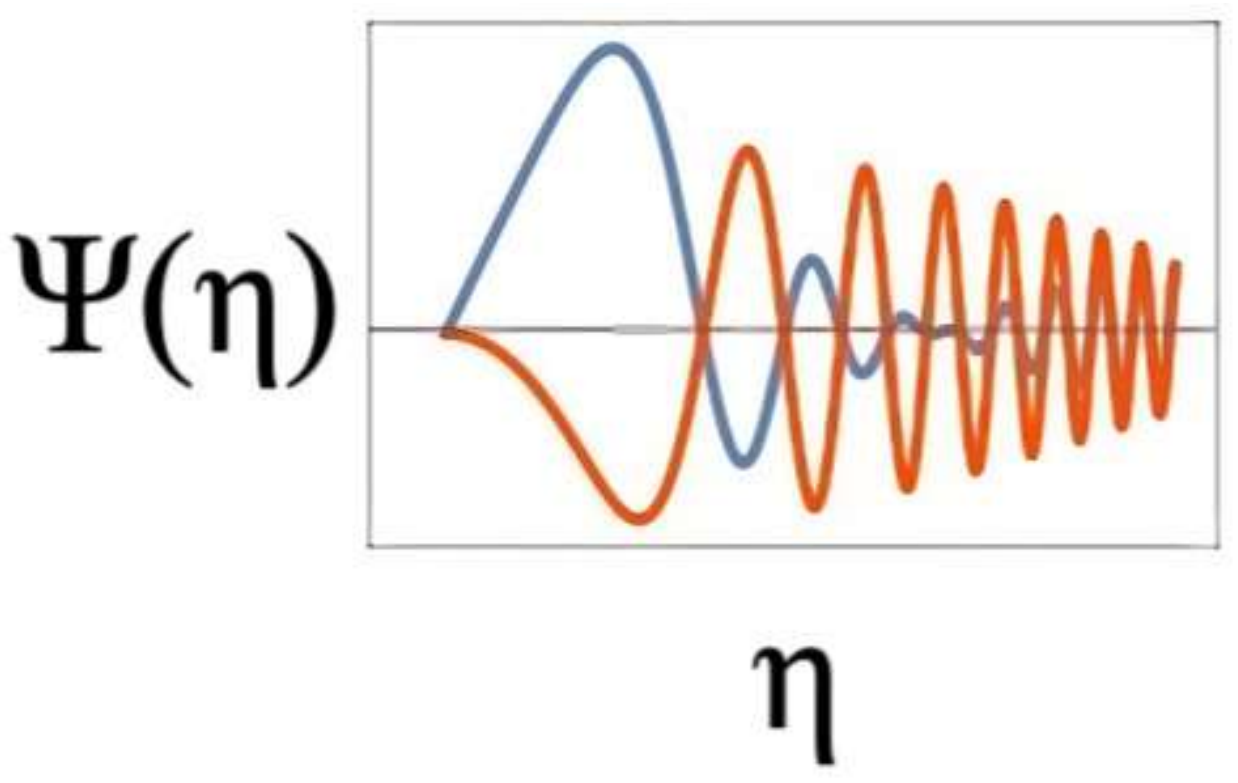}
    \caption{On the upper-left, an Argand-type diagram of the cut distribution of real and imaginary numerical sample family solutions of the wave equation (\ref{psieta+}) for the wave-function \(\Psi(\eta)\) assuming the naturalness condition for $|\gamma| = 1$. 
    On the upper-right and center-below plots of sample individual solutions. The upper-left image corresponds to sampling the boundary conditions 
    $\Psi(1)$ and $\Psi'(1)$. The upper-right image corresponds to the boundary conditions  $\Psi(1) =1$ and $\Psi'(1) =0$, while the center-below image to the boundary conditions  $\Psi(1) =0$ and $\Psi'(1) =1$.}
  \label{reWFpsieta}
\end{figure*}
\begin{figure*}[htbp]
   \centering
\includegraphics[width=0.49\textwidth]{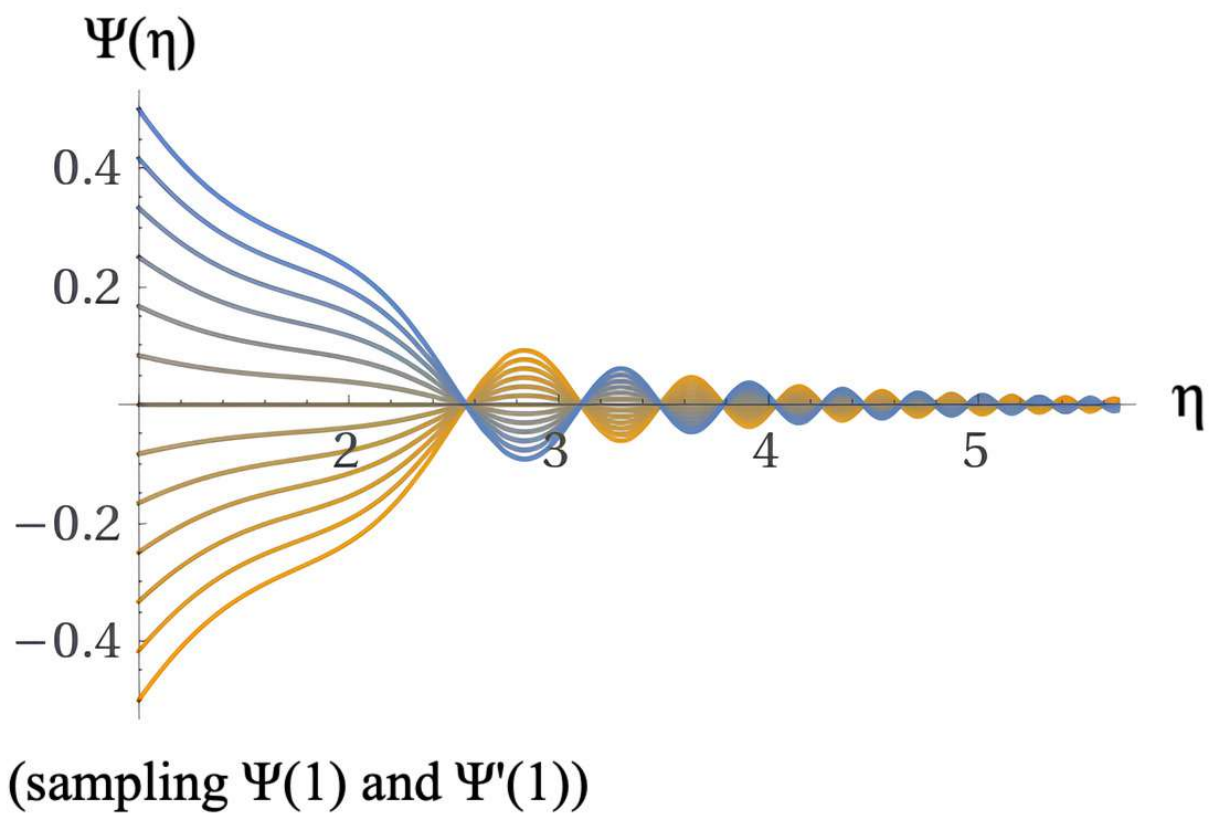}
\includegraphics[width=0.46\textwidth]{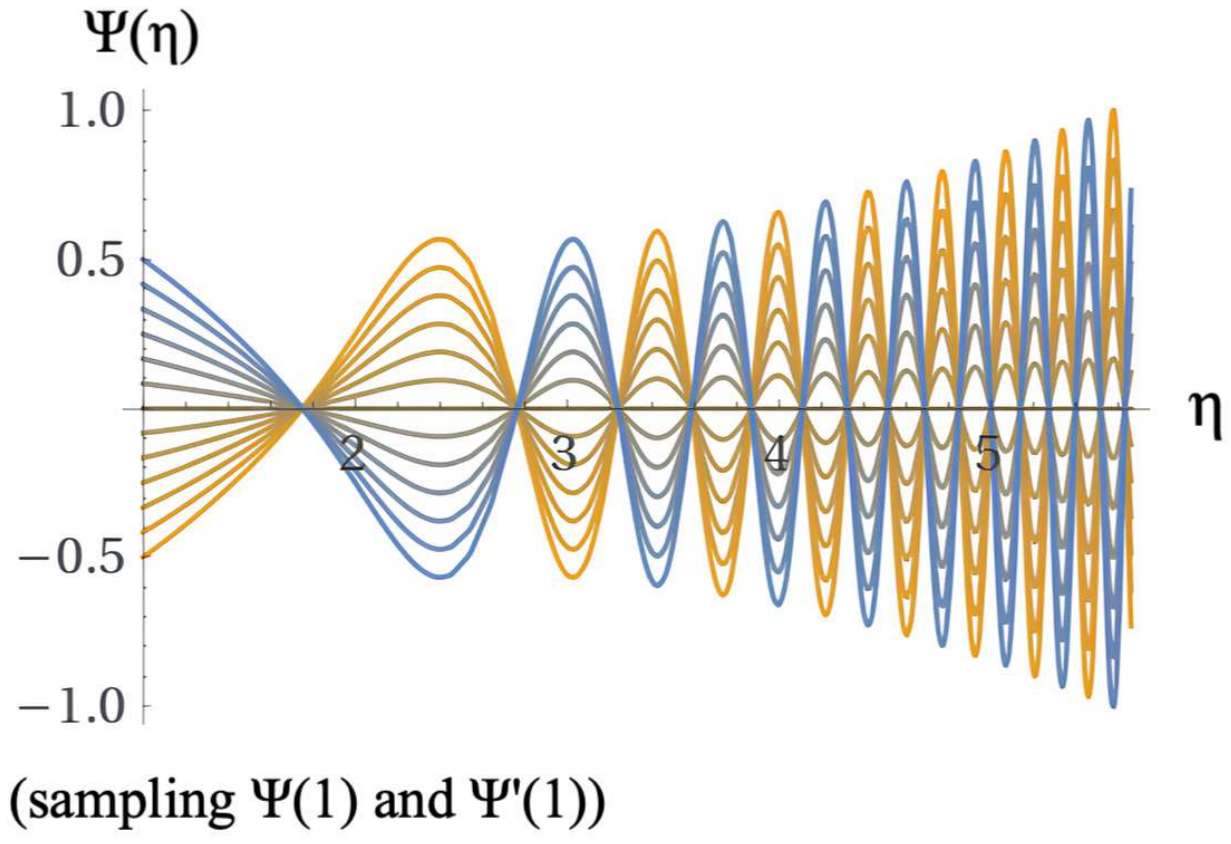}\\
    \caption{Sample solutions of the wave equation (\ref{psieta++}) for the wave-function  \(\Psi(\eta)\) assuming $\gamma = i |\gamma|$ and the naturalness condition. 
    The left figure corresponds to $|\gamma| = -1$, while on the right $|\gamma| = 1$. }
  \label{imWFpsieta*+}
\end{figure*}
Figure~\ref{reWFpsieta} shows on the upper-left image an Argand-type diagram which refers to the transversal structure of the distribution of real and imaginary family sample  solutions of the wave equation (\ref{psieta+}) for the  wave-function $\Psi(\eta)$, depicting its geometric properties.
The diagram indicates an overlapping winding between the real and imaginary components of the multi-valued function $\Psi(\eta)$, reflecting the intense correlation between matter and energy encapsulated in the potential ${\cal V}(\eta)$. Inserting a complex polar representation of $\Psi(\eta) \equiv \sqrt{r}e^{i\theta}$, a complete description of this wave-function is portrayed by copies, or branch-cut Riemann sheets, in the complete complex cut-plane
extending from $-\infty \leq \theta \leq \infty$, forming a Riemann surface. On the upper-right and below-center, images of $\Psi(\eta)$ as a function of $\eta$. The figures exhibit a similar wave behavior, given the parametric choices, particularly concerning the lower central figure to the historical result obtained by
J.B. Hartle and S.W. Hawking~\cite{Hartle1983}. According to the authors, the solutions would indicate the de Sitter space expands without limit, and in our view evolving towards a state of harmony.
The behavior of the solutions indicates that the complex $\Psi(\eta)$ function is meromorphic, --- that is holomorphic and therefore analytic on the upper and below cut planes.
Figure~\ref{imWFpsieta*+} show sample solutions of the wave equation (\ref{psieta++}) for the wave-function \(\Psi(\eta)\),  assuming the naturalness condition, and $\gamma = i |\gamma|$,   with $|\gamma| = -1$ on the left images of the figure and $|\gamma| = 1$ on the right images of the figure.
The result on the left of figure~\ref{imWFpsieta*+} resembles the predictions of the big bounce theory, which is based on the notion that the evolution of the universe corresponds to a cyclical and non-linear natural phenomenon that sustains a deceleration stage of the expansion process conforming a contraction phase.
The results of the images on the right of figure \ref{imWFpsieta*+} show an expanding and cyclical universe, with the amplitudes of the $\Psi(\eta)$ wave-function 
growing systematically intensely in contrast to increasingly shorter Planck time intervals.
This behavior characterizes a universe in accelerated expansion as a result of the reconfiguration of primordial matter and energy 
and the capture of small- and large-scales of spacetime
due to the non-commutative symplectic algebraic structure. 
\begin{figure*}[tbhp]
\centering
\includegraphics[width=0.46\textwidth]{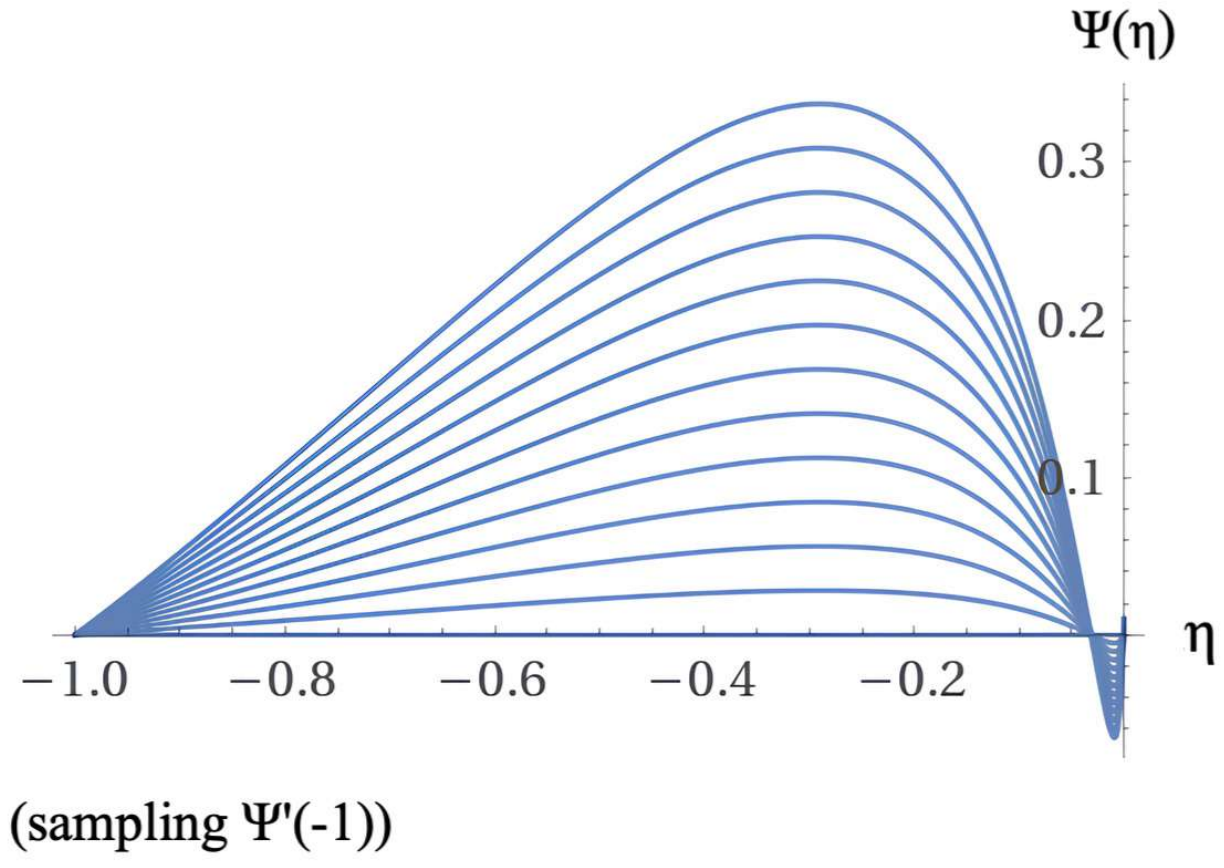}
\includegraphics[width=0.47\textwidth]{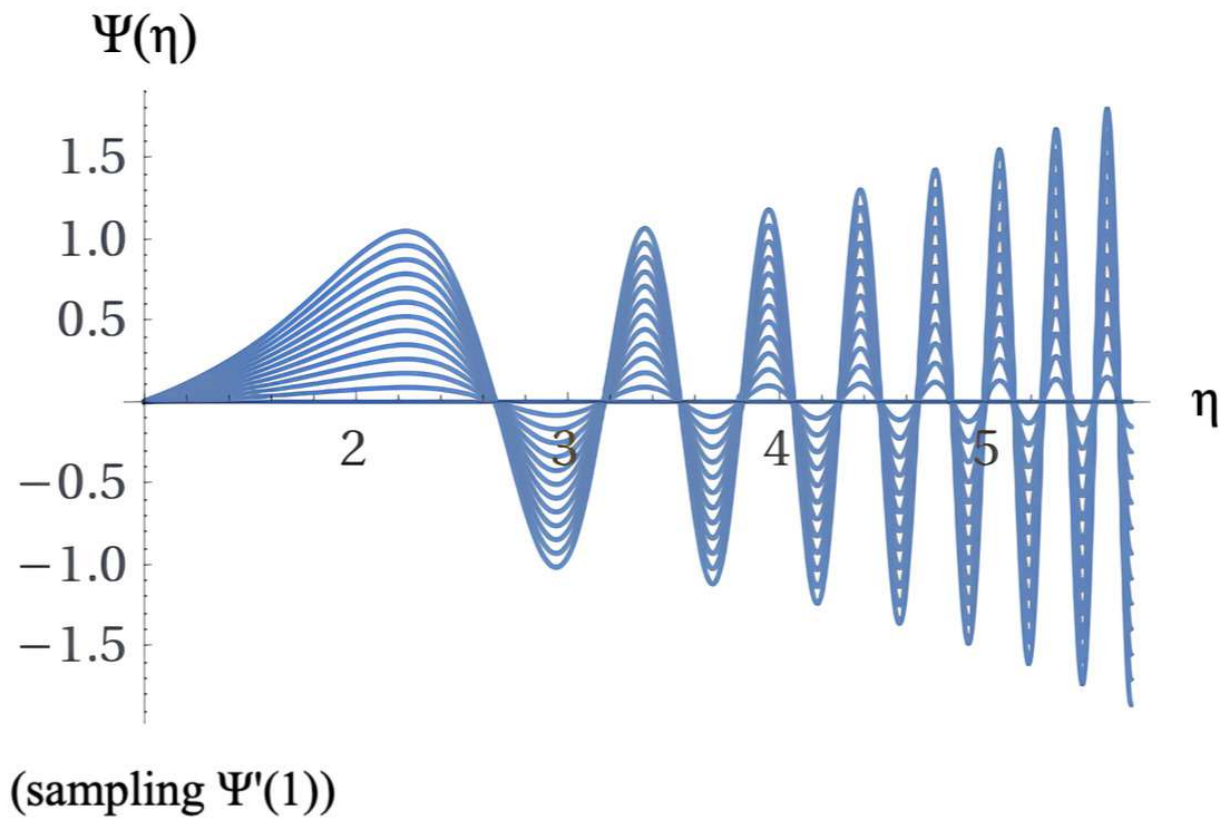}\\
\includegraphics[width=0.46\textwidth]{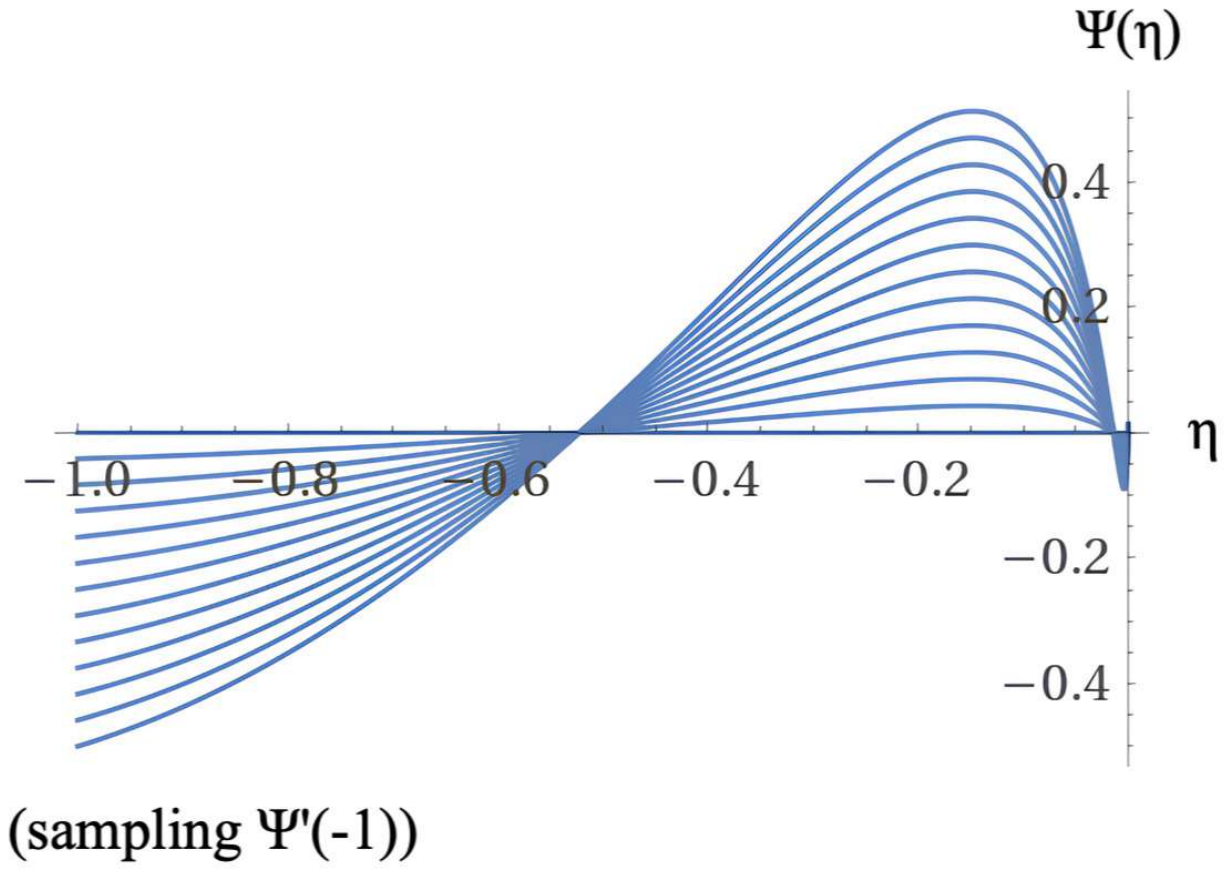}
\includegraphics[width=0.47\textwidth]{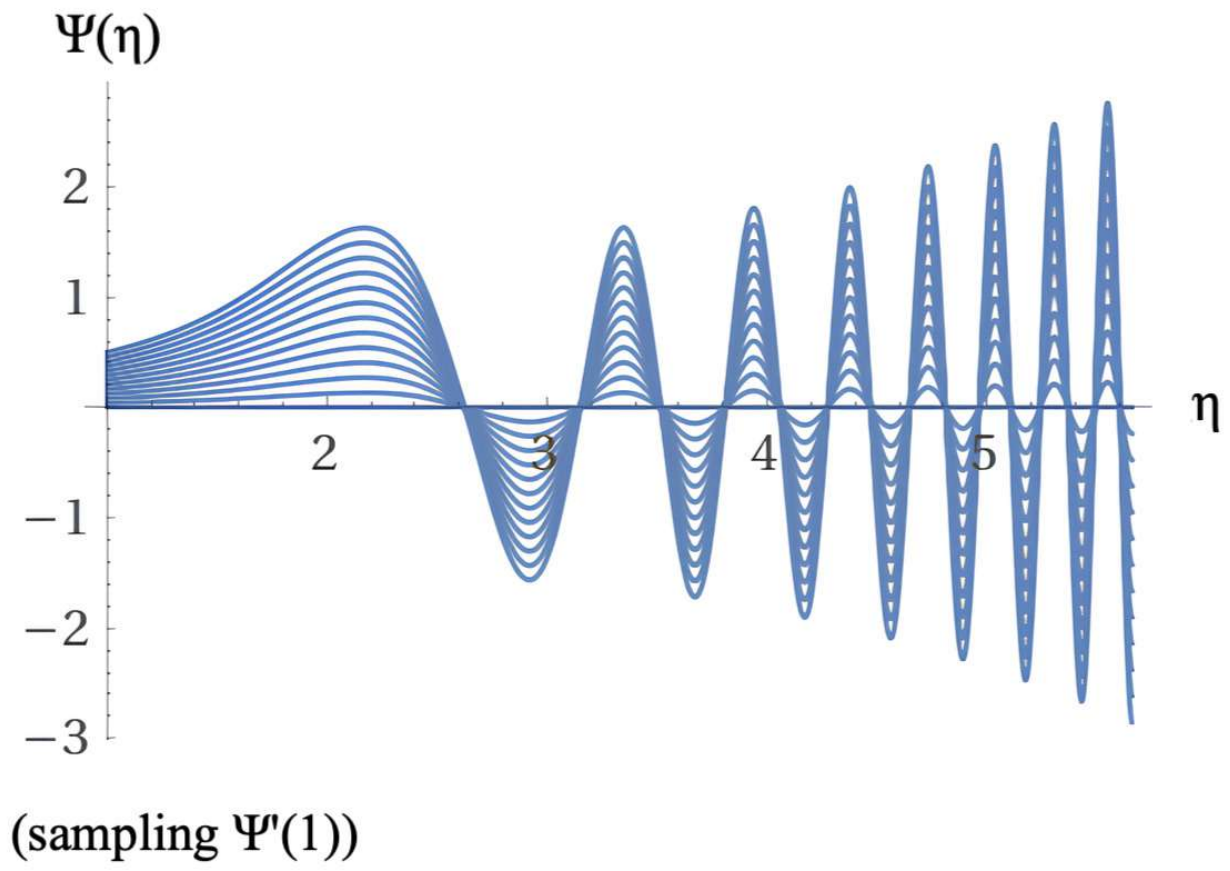}\\
\caption{Sample family solutions of equation (\ref{psieta++}) for the wave-function $\Psi(\eta)$ assuming the naturalness condition. The upper-left and lower-left figures correspond respectively to the boundary conditions $\Psi(-1)=0$ and $\Psi(-1) = -1$. The upper-right and lower-right figures correspond respectively to the boundary conditions  $\Psi(1)=0$ and $\Psi(1) = 1$.} \label{NCWFU}
\end{figure*} 
Figure~\ref{NCWFU} show sample family solutions of equation (\ref{psieta++}) for the wave-function $\Psi(\eta)$ assuming the naturalness condition. The upper-left and lower-left figures correspond respectively to the boundary conditions $\Psi(-1)=0$ and $\Psi(-1) = -1$ while the upper-right and lower-right figures correspond respectively to the boundary conditions  $\Psi(1)=0$ and $\Psi(1) = 1$.
According to BCQG, the transition between the mirror universe, in a contracting phase, and its current expanding counterpart occurs through a quantum leap, exemplified in the figure~\ref{NCWFU}, corresponding to the upper-right and lower-right images and boundary conditions $\Psi(1)=0$ and $\Psi(1) = 1$.
\subsubsection{$\Psi(\xi)$ solutions}

The algebraic solution for the differential equation (\ref{psixi}) is 
\begin{equation}
    \Psi(\xi) = c_1 e^{i\xi/2}Ai\Bigl(\frac{4 \xi -3}{4 \sqrt[3]{3}} \Bigr) + c_2 e^{i\xi/2}Bi\Bigl(\frac{4 \xi -3}{4 \sqrt[3]{3}} \Bigr). \label{asPsixi}
\end{equation}

\begin{figure*}[htbp]
   \centering
\includegraphics[width=0.51\textwidth]{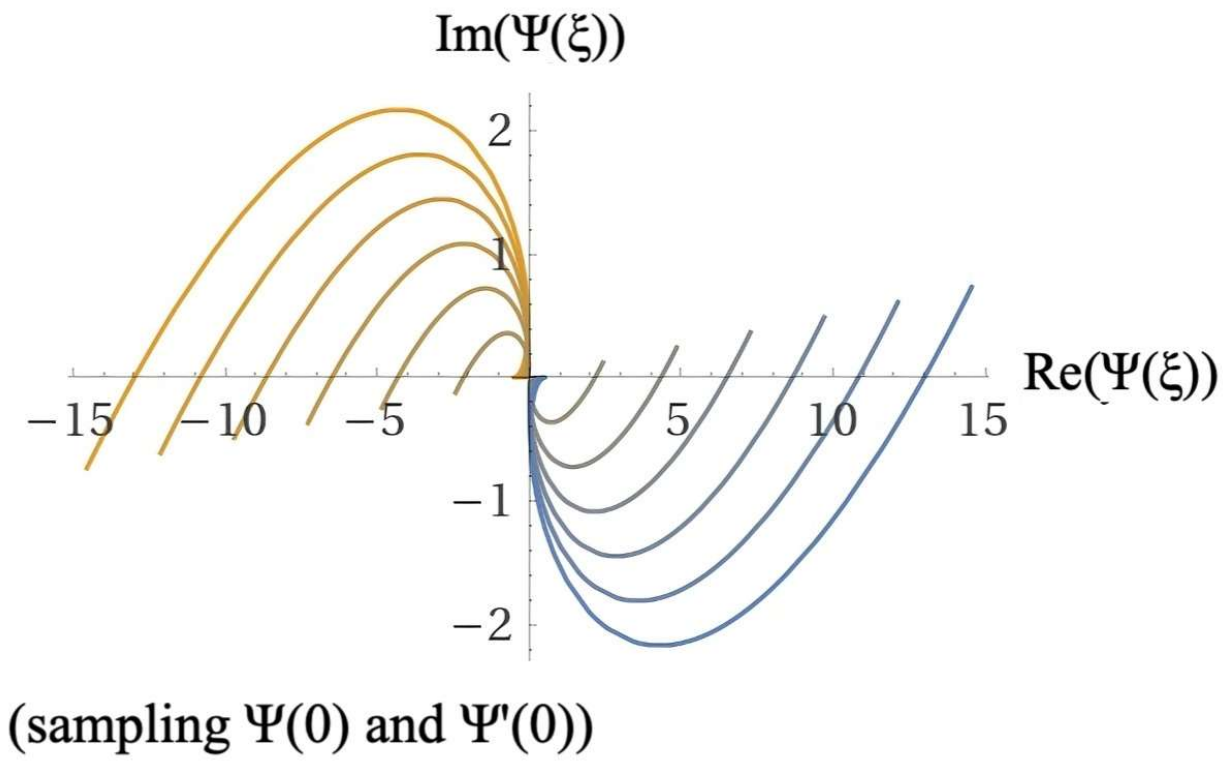}
\includegraphics[width=0.48\textwidth]{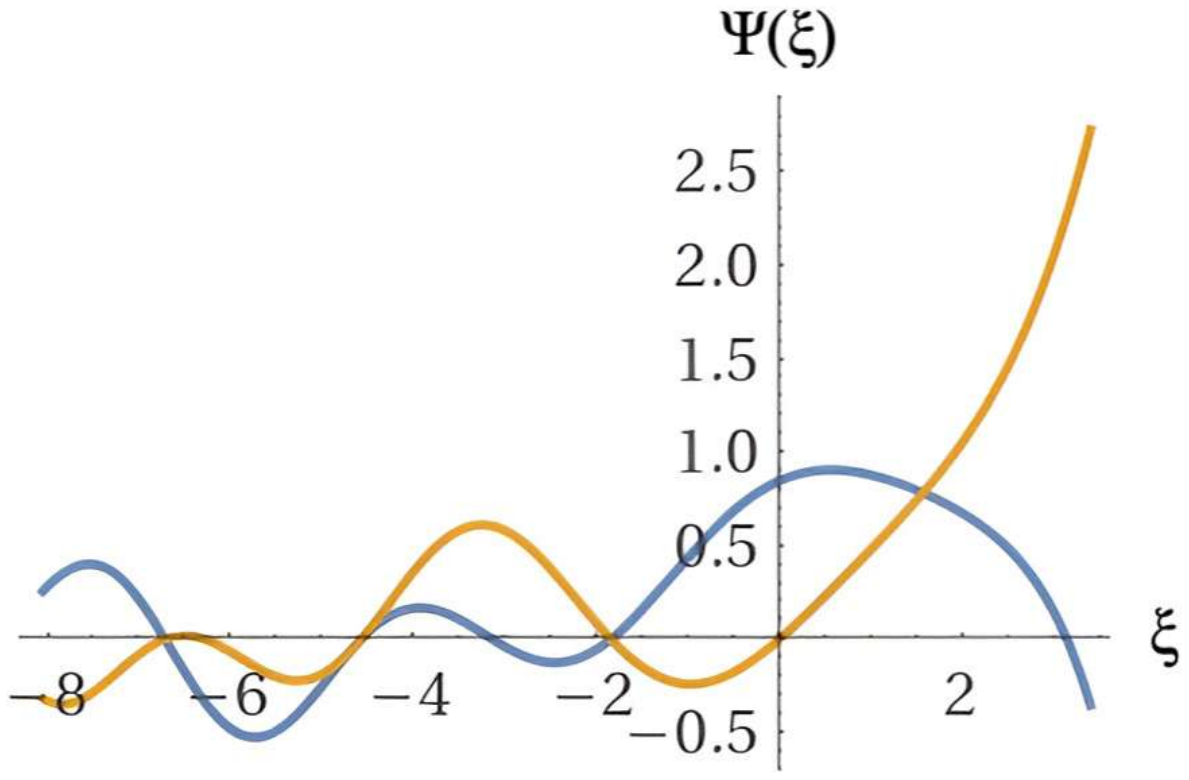}
    \caption{Sample family solutions of the wave equation (\ref{psixi}) for the wave-function  $\Psi(\xi)$, assuming the naturalness condition. 
    The figure on the left shows an Argand-type plot of the cut distribution of real and imaginary numerical sample solutions of the  wave-function $\Psi(\xi)$. The figure on the right show plots of the solution of equation (\ref{psixi}) given in (\ref{asPsixi}). The blue lines correspond to real parts while orange lines to imaginary components.}
  \label{WFpsixi}
\end{figure*}
Figure \ref{WFpsixi} exhibit sample family solutions of the wave equation (\ref{psixi}) for the wave-function  $\Psi(\xi)$, assuming the naturalness condition. 
    The figure on the left shows an Argand-type transversal structure mapping the distribution of real and imaginary numerical sample solutions of the  wave-function $\Psi(\xi)$. The figure on the right show plots of the solution of equation (\ref{psixi}) given in (\ref{asPsixi}). The blue lines correspond to real parts of the solutions while the orange lines to imaginary components. The behavior of the solutions corresponding to the wave function $\Psi(\xi)$, as a function of $\xi(\tau)$, a support field that is both dual and complementary to the scale factor $\eta(\tau)$, once again highlights a singular aspect of BCQG, the emanation of a mirror universe, nested within ours in the space-time domain of negative cosmological time, offering a consistent alternative to the original proposal of the inflation model of a universe arising from nothing as a result of vacuum fluctuations.
    Due to the nature of the wave equation of the dual field $\xi$, the Argand-type diagram reveals that the imaginary part of the solutions is restricted to the negative sector of the real part, while the opposite occurs for the real part. A crucial aspect of the solutions, highlighted in the image on the right of the figure \ref{WFpsixi} is their continuity, indicating a hypothetical connection between the current universe and its mirror counterpart BCQG.
\subsubsection{$\Psi(\varphi)$ solutions}
\begin{figure*}[htbp]
   \centering
\includegraphics[width=0.50\textwidth]{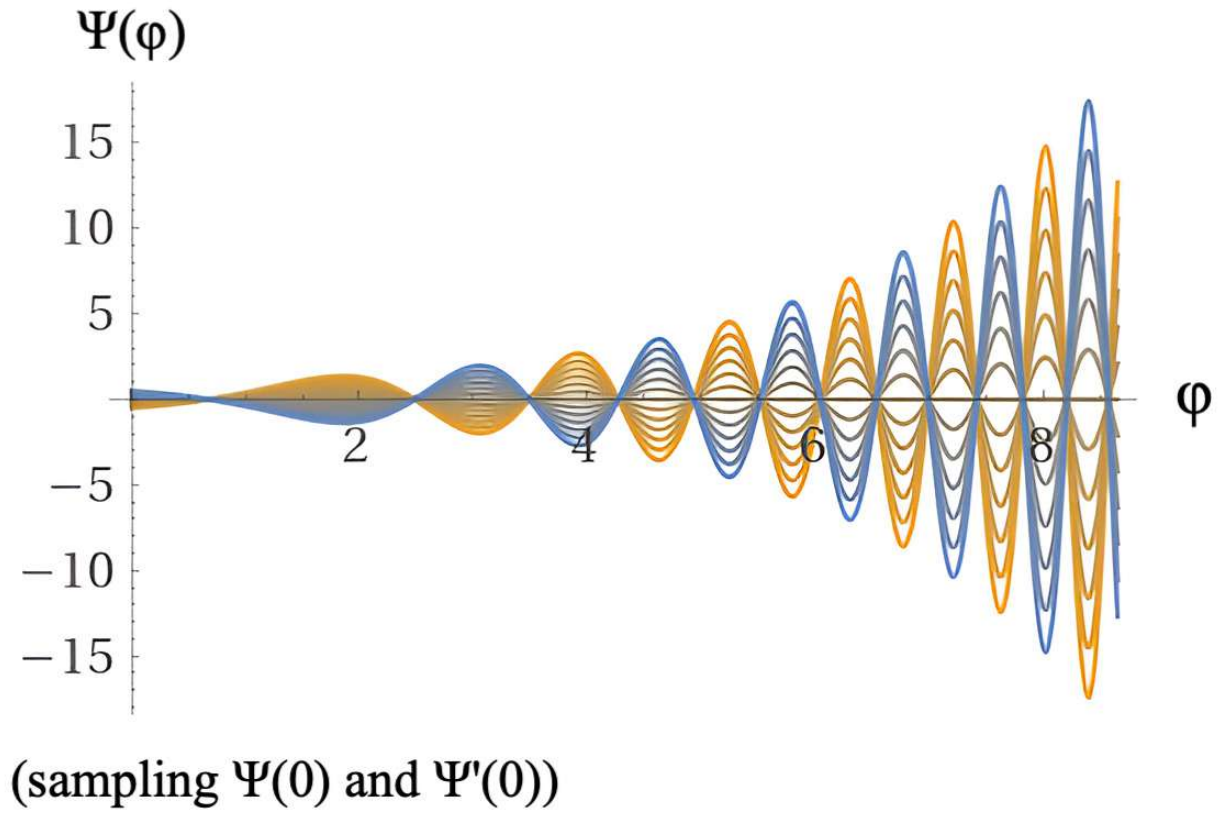}
\includegraphics[width=0.49\textwidth]{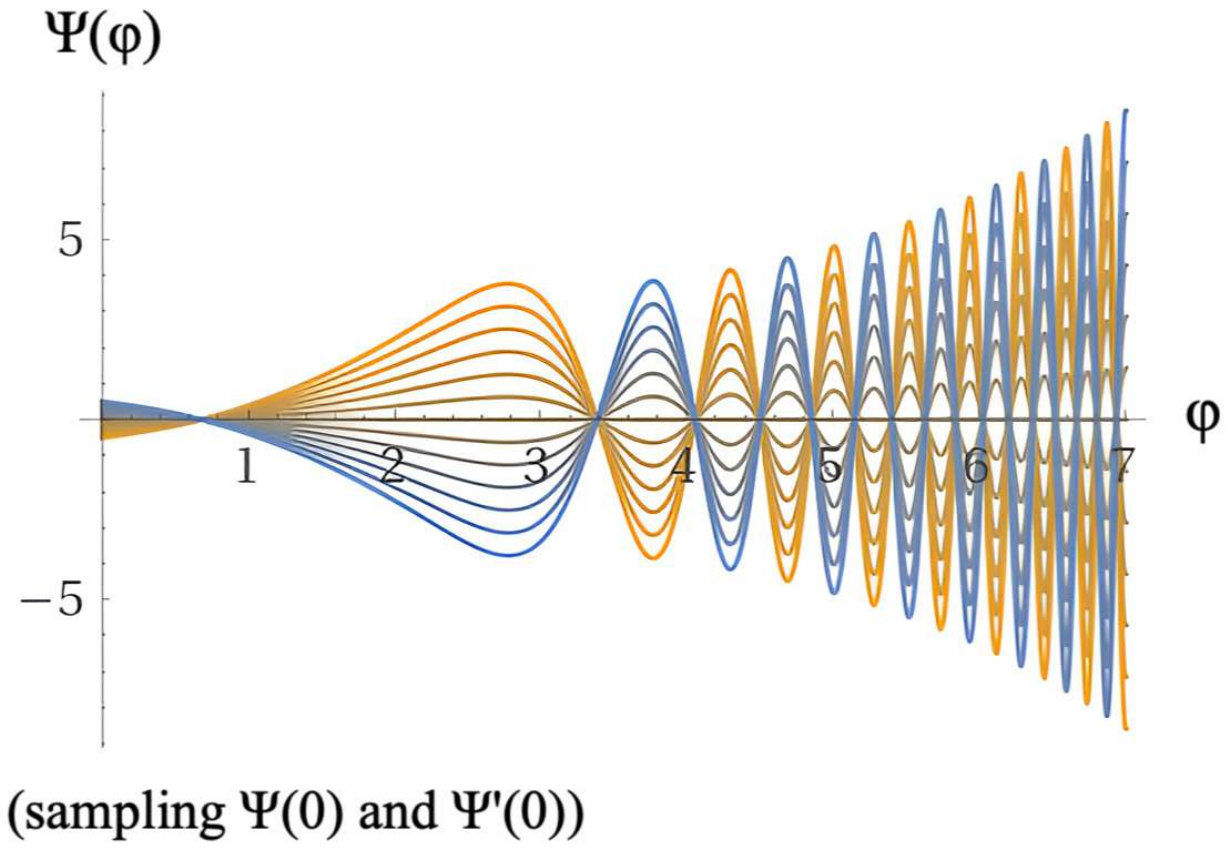}
    \caption{Sample family solutions of the wave equation (\ref{psivarphi++}) for the wave-function  $\Psi(\varphi)$, assuming the naturalness condition. 
     The figure on the left corresponds to chaotic inflation while the figure on the right to non-chaotic inflation.}
  \label{WFpsivarphi2}
\end{figure*}

The algebraic solution for the differential equation (\ref{psivarphi}), in case of chaotic inflation, is
\begin{eqnarray}
    \Psi(\varphi) & = & c_1 e^{-\frac{1}{6}i\varphi(3\varphi+4)}H_{-\frac{1}{2}-\frac{i}{9}}\Biggl(\frac{1}{6} \sqrt[4]{-1} (6 \varphi + 1) \Biggr) \nonumber \\
   && + c_2 e^{-\frac{1}{6}i\varphi(3\varphi+4)} ~_1F_1 \Biggl(\frac{1}{4} + \frac{i}{18}; \frac{1}{2}; \frac{1}{36}i (6 \varphi + 1)^2  \Biggr). 
\end{eqnarray}
\begin{figure*}[htbp]
   \centering
\includegraphics[width=0.49\textwidth]{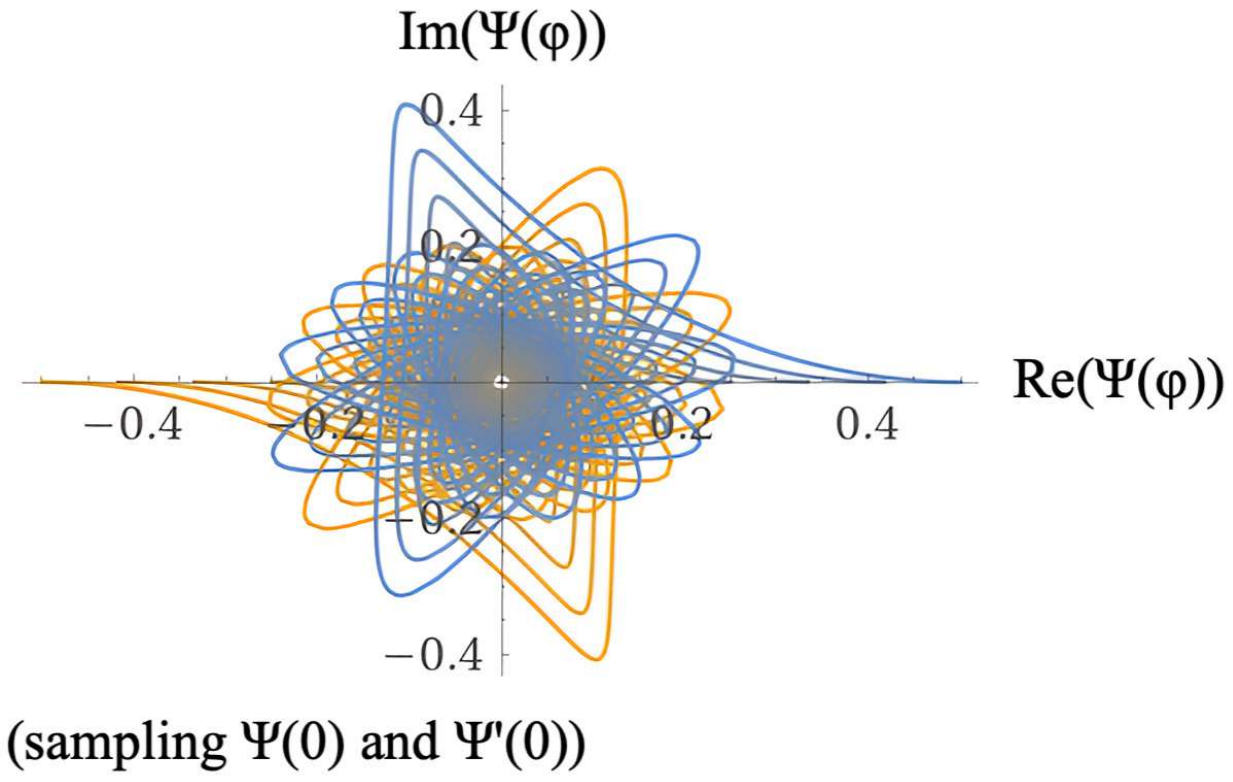}
\includegraphics[width=0.50\textwidth]{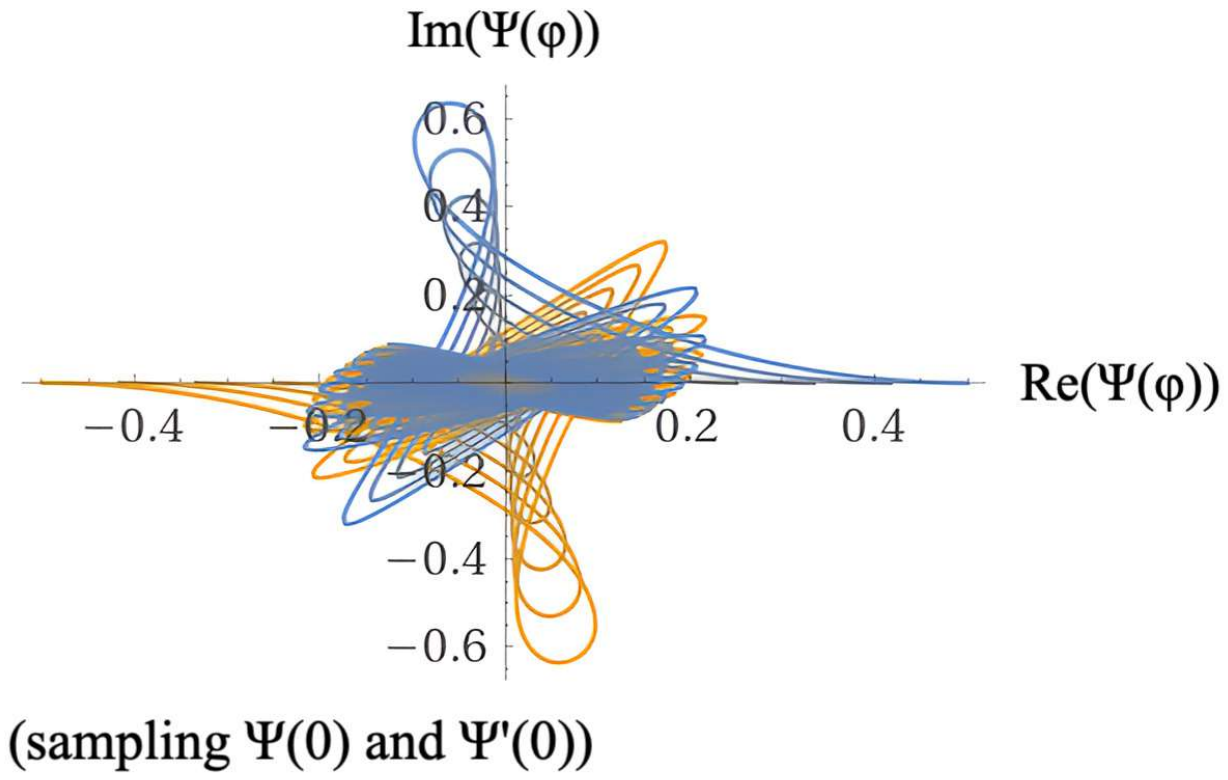}
    \caption{Sample family solutions of the wave equation (\ref{psivarphi+}) for the wave-function  $\Psi(\varphi)$, assuming the naturalness condition. 
    The figures show Argand-type plots of real and imaginary numerical sample solutions of the  wave-function  $\Psi(\varphi)$. The figure on the left corresponds to chaotic inflation while the figure on the right to non-chaotic inflation.}
  \label{WFpsivarphi}
\end{figure*}
Figure \ref{WFpsivarphi} show sample family solutions of the wave equation (\ref{psivarphi+}) for the wave-function  $\Psi(\varphi)$, assuming the naturalness condition. 
    Both figures show Argand-type plots of real and imaginary numerical sample solutions of the  wave-function $\Psi(\varphi)$. The figure on the left corresponds to chaotic inflation while the figure on the right to non-chaotic inflation. 
The analyses previously presented regarding Argand-type plots of real and imaginary numerical sample solutions of the  wave-function 
$\Psi(\eta)$ apply similarly to the present case of the wave function $\Psi(\varphi)$.
Figure \ref{WFpsivarphi2} show sample family solutions of the wave equation (\ref{psivarphi++}) for the wave-function  $\Psi(\varphi)$, assuming the naturalness condition. 
      The results for the wave-function $\Psi(\eta)$ corresponding to figures \ref{imWFpsieta*+} and \ref{NCWFU}, on the right, when compared to the results 
     of figure \ref{WFpsivarphi2}, for the wave-function $\Psi(\varphi)$ present unique and consistent characteristics and similarities.
     The evolutionary behavior for both solutions, indicates a cyclical and rapidly expanding universe, with the amplitudes of both wave functions,  $\Psi(\eta)$ and $\Psi(\xi)$, systematically increasing in contrast to the systematic reduction of the corresponding Planck time intervals, a compelling indication of cosmic acceleration in the inflationary period. 
       To our knowledge, this is the first time that solutions of a wave equation of an inflaton field are known, in the context of quantum gravity, evidencing related and complementary behavior of the inflaton field and the cosmic scale factor. These results further indicate that the conventional inflation model, although introduced historically in an ad hoc manner, includes crucial elements that characterize the cosmic acceleration drive. 
     This behaviour indicates a growing disruptive evolution increase in the branched-gravitation expansion phase, prior to the BCQG transition region characterized by the overcoming of the primordial singularity, as predicted in the standard model, and prior the wave-contraction phase. 
     Furthermore, as we will see later, when we consider the dynamical equations involving the cosmic scale factor, the presence of the inflaton field in the present formulation of BCQG adheres consistently to the reconfiguration of spacetime as a result of a non-commutative symplectic algebraic structure, contributing in a unique manner for cosmic acceleration.
     The branched transition region may be outlined according to three different perspectives: (i) a classical view of a kind of ‘transition topological portal’, (ii) a quantum view region which contemplates a topological quantum leap, and (iii) a mixed conception in which both previous conceptions consistently intersect, shaped on the basis of the Bekenstein criterion (see~\cite{Zen2024}). The topologically foliated branch-cut quantum structural representation of the transition region involving the contraction and expansion phases resembles a topological spacetime shortcut, as a kind of foliated wormhole structure. Theoretical implications of the existence of topological shortcuts in spacetime imply a challenge to our understanding of fundamental physical principles, such as causality or still open topics such as the origin of primordial cosmological material seeds. Accordingly, the standard inflation model conception of the creation of the present universe from nothing, as a result of virtual vacuum fluctuations, falls apart when we examine the results of figures \ref{NCWFU} and \ref{WFpsixi}. These results indicate that the present universe might have its origin in an earlier phase through a possible distant spacetime shortcut. We may conjecture if this tiny correction would be, in principle, detectable by homodyne-type measurements, --- a method of extracting information encoded as modulation of the phase and/or frequency of an oscillating signal, a gravitational wave for instance. Additionally, we may even conjecture if this tiny correction would be detectable  
     after long propagation lengths for a wide range of throat radii and distances to the shortcut, even if the detection takes place very far away from the throat, where the spacetime is very close to a flat geometry.

\section{Final remarks and conclusion}

Quantum field theory in non-commutative spacetime leads to an uncertainty relation for coordinates analogous to the Heisenberg uncertainty principle, whose lower bound has given rise to the minimum scale problem. An unique feature of non-commutative field theory 
is the realization of a mixing between ultraviolet and infrared radiation (UV/IR), which characterizes an interrelation between short and long-range scales, absent in commutative quantum field theories (see for instance~\cite{Kontsevich}). 
The theoretical problems above were also identified by Seiberg–Witten~\cite{SeibergWitten}.
On the basis of the weak gravity conjecture, the notion of hierarchical UV/IR mixing was implemented with scalar fields by L\"ust and Palti~\cite{Lust}.
In addition, the deep-low IR and far-high UV connection, according to Craig and  Koren, by studying IR dynamics from UV divergences, in the context of UV/IR mixing, could even solve (part of) the hierarchical naturalness problem~\cite{Craig}. 

Different hypotheses can be raised to explain the cosmic inflation acceleration. A compelling possibility emulates from the electroweak  hierarchy problem, based on 
an inspiring article~\cite{Minwalla1}. The authors, in their study of perturbative dynamics of a particular set of non-commutative field theories, found an intriguing mixing of the ultraviolet (UV) and the infrared (IR) domains, identified with the observation that high energy UV virtual particles in loops produce non-analyticity at low momentum, interpreted as IR divergences. This finding, a realization of the mixing between the the short-range, high-energy ultraviolet (UV) and long-range, low-energy infrared (IR) phenomena, according to the authors, arises from the underlying non-commutativity, a phenomenon reminiscent of the channel duality of the double twist diagram in open string theory  (for the details, see~\cite{Minwalla1}). (For an alternative discussion see~\cite{Sabido}).
In short, the prevailing conception in field theory is that the
 origin of the IR/UR mixture results from the capture of both large
 and small scales by the presence of a non-commutative structure.
Translating this conception to quantum gravity,
the non-commutative symplectic quantum structure of spacetime, through the capture of small-, medium- and large-scales, generates, on the one hand, a topological restructuring of space-time and, on the other, a restructuring of the primordial distribution of matter and energy.

\section{Acknowledgements}
P.O.H. acknowledges financial support from PAPIIT-DGAPA (IN116824). F.W. is supported by the U.S. National Science Foundation under Grant PHY-2012152.

\section*{Author Contributions}
Conceptualization, C.A.Z.V.; methodology, C.A.Z.V. and B.A.L.B. and P.O.H and J.A.deF.P. and D.H. and F.W. and M.M.; software, C.A.Z.V. and B.A.L.B. and M.R. and M.M.; validation, C.A.Z.V. and B.A.L.B. and D.H. and P.O.H. and J.A.deF.P. and F.W.; formal analysis, C.A.Z.V. and B.A.L.B. and P.O.H. and J.A.deF.P. and D.H. and F.W.; investigation, C.A.Z.V. and B.A.L.B. and P.O.H. and J.A.deF.P. and M.R. and M.M. and F.W.; resources, C.A.Z.V.; data curation, C.A.Z.V. and B.A.L.B.; writing—original draft preparation, C.A.Z.V.; writing—review and editing, C.A.Z.V. and B.A.L.B. and P.O.H. and J.A.deF.P. and D.H. and M.R. and M.M. and F.W.; visualization, C.A.Z.V. and B.A.L.B.; supervision, C.A.Z.V.; project administration, C.A.Z.V.; funding acquisition (no funding acquisition). All authors have read and agreed to the published version of the manuscript.

\appendix

\section{A three-field Faddeev–Jackiw symplectic algebra extension}\label{A}
From equation (\ref{eq-6}), 
the following conditions for the matrix elements $M_{ij}$ of the $M$ matrix apply:
\begin{eqnarray}
u & = & M_{11} {\tilde u} + M_{12} {\tilde p}_u + M_{13} {\tilde v} + M_{14} {\tilde p}_v 
+ M_{15} {\tilde \phi} + M_{16} {\tilde p}_\phi; 
\nonumber \\ 
p_u & = & M_{21} {\tilde u} + M_{22} {\tilde p}_u + M_{23} {\tilde v} + M_{24} {\tilde p}_v 
+ M_{25} {\tilde \phi} + M_{26} {\tilde p}_\phi; 
\nonumber \\ 
v & = & M_{31} {\tilde u} + M_{32} {\tilde p}_u + M_{33} {\tilde v} + M_{34} {\tilde p}_v 
+ M_{35} {\tilde \phi} + M_{36} {\tilde p}_\phi; 
\nonumber \\ 
p_v & = & M_{41} {\tilde u} + M_{42} {\tilde p}_u + M_{43} {\tilde v} + M_{44} {\tilde p}_v 
+ M_{45} {\tilde \phi} + M_{46} {\tilde p}_\phi; 
\nonumber \\
\phi & = & M_{51} {\tilde u} + M_{52} {\tilde p}_u + M_{53} {\tilde v} + M_{54} {\tilde p}_v 
+ M_{55} {\tilde \phi} + M_{56} {\tilde p}_\phi; 
\nonumber \\ 
p_\phi & = & M_{61} {\tilde u} + M_{62} {\tilde p}_u + M_{63} {\tilde v} + M_{64} {\tilde p}_v 
+ M_{65} {\tilde \phi} + M_{66} {\tilde p}_\phi.
\label{eq-7}
\end{eqnarray}

Adopting as a simplification the condition $\tilde{u}  = u$, $\tilde{v} = v$, and $\tilde{\phi} = \phi$, since this setting does not change the structure of the approach, maintaining the same polynomial orderings of the corresponding variables, - and as a consequence the same structure of the matter and energy potential  in the BCQG action, we can reduce the matrix $M$ to the expression
\begin{eqnarray}
\left( M\right)  & = &
\left(
\begin{array}{cccccc}
1 & 0 & 0 & 0 & 0 & 0 \\
M_{21} & M_{22} & M_{23} & M_{24}& M_{25} & M_{26}  \\
0 & 0 & 1 & 0 & 0 & 0 \\
M_{41} & M_{42} & M_{43} & M_{44} & M_{45} & M_{46} \\
0 & 0 & 0 & 0 & 1 & 0 \\
M_{61} & M_{62} & M_{63} & M_{64} & M_{65} & M_{66} \\
\end{array}
\right).
\label{eq-8}
\end{eqnarray}
On the basis on the sets (\ref{eq-5a}) and (\ref{eq-5b}), we consider first the Poisson brackets that involve only one-coordinate and one-momentum components, which will lead to a first level of symplectic conditions to determine the matrix elements $M_{ik}$. Then, we consider the Poisson brackets of two momenta components, whose expressions conditions are more involved, which will lead to a second level of symplectic conditions to determine the remaining matrix elements $M_{ik}$. The Poisson brackets, involving only one coordinate with one momentum, yield:
\begin{eqnarray}
\{ u, p_u \} & = & M_{22} = 1; \quad 
\{ u, p_v \} = M_{42} = \gamma; \quad  
\{ u, p_\phi \} =  M_{62} = \beta; 
\nonumber \\ 
\{ v, p_u \} & = & M_{24} = \chi; \quad
\{ v, p_v \} = M_{44} = 1; \quad 
\{ v, p_\phi \} = M_{64} = \zeta; 
\nonumber \\ 
\{ \phi , p_u \} & = & M_{26} = \delta; \quad 
\{ \phi , p_v \} = M_{46} =-\varsigma; \quad  
\{ \phi , p_\phi \} = M_{66} = 1.
\label{eq-8-0}
\end{eqnarray}

The Poisson brackets, with one coordinate and one momentum each,
lead to
\begin{eqnarray}
\{ p_u,p_v\} & = &  \gamma M_{21} - M_{41} + M_{23} - \chi M_{43}
+ \varsigma M_{25} - \delta M_{45} = \alpha_1; 
\nonumber \\
\{ p_u,p_\phi\} & = & \beta M_{21} - M_{61} + \zeta M_{23}
-\chi M_{63} + M_{25} - \delta M_{65} = \alpha_2;
\nonumber \\
\{ p_v,p_\phi \} & = & \beta M_{41} - \gamma M_{61} + \zeta M_{43}
- M_{63} + M_{45} - \varsigma M_{65} = \alpha_3.
\label{eq-8b}
\end{eqnarray}

\section{Canonical form of the two-fields non-commutative Hamiltonian}\label{B}

Equation (\ref{HTinverted+++}) can be rewritten as 
\begin{eqnarray}
 {\cal H} \Psi(u,v,\phi) 
       & \! = \! &  \! \Biggl[ \! \Bigl( \! {\cal C}_1  \frac{\partial^2}{\partial u^2} \! + {\cal C}_2  \frac{\partial^2}{\partial v^2}\! +  {\cal C}_3  \frac{\partial^2}{\partial \phi^2} \! +  {\cal C}_4  \frac{\partial}{\partial u}\frac{\partial}{\partial v}\!+ 
        {\cal C}_5 
       \frac{\partial}{\partial u}\frac{\partial}{\partial \phi} \! +  {\cal C}_6 \! \frac{\partial}{\partial v}\frac{\partial}{\partial \phi} \Bigr) \nonumber \\
&&       +  \frac{1}{u^{3 \alpha-1}} \Bigl( i\gamma \frac{\partial}{\partial u} + \alpha {u} - i \frac{\partial}{\partial v}  - \alpha {v} + i\varsigma \frac{\partial}{\partial \phi} + \alpha \phi \Bigr)   
      + 2 V(\phi)  \nonumber \\
&&
       + \Bigl( \! g_r \! - g_m {u} -  g_k {u}^2 \! - g_q {u}^3 +     
 g_{\Lambda} {u}^4 
   \! + \frac{g_s}{{u}^2}  \! \Bigr) \! \Biggr] \! \Psi(u,v,\phi) = 0  , \nonumber \\
 & &\longrightarrow  
    \Biggl[\, \, \,\sum_{i,j=u,v,\phi}  a_{ij}({\bm{\psi}}) \, \frac{\partial^2}{\partial \varphi_i \partial \varphi_j}    + 2 V(\phi)  \label{HTinverted*} \\
 & &
      + \frac{1}{u^{3 \alpha-1}} \Bigl( i\gamma \frac{\partial}{\partial u}  + \alpha {u}  - i \frac{\partial}{\partial v}   - \alpha {v}  +  i\varsigma \frac{\partial}
{\partial \phi} + \alpha \phi \Bigr)   
      \nonumber \\
&&  
       + \Bigl(  g_r  - g_m {u} -  g_k {u}^2 \! - g_q {u}^3 +     
 g_{\Lambda} {u}^4 
   \! + \frac{g_s}{{u}^2}  \Bigr)  \Biggr]  \Psi(u,v,\phi) = 0  , \nonumber 
\end{eqnarray} 
with the relation between the $a(\mathbf{\phi})_{ij}$ and the ${\cal C}_i$ coefficients defined in table \ref{tableac}.
\begin{table}[htbp]
    \centering
    \begin{tabular}{|c|c|c|c|c|c|} \hline
        $a_{uu}(\bm{\psi})$ & $a_{vv}(\bm{\psi})$ &  $a_{\phi \phi}(\bm{\psi})$ & $a_{uv}(\bm{\psi})$ & $a_{u \phi}(\bm{\psi})$ &  $a_{v \phi}(\bm{\psi})$ \\ 
        \hline
            ${\cal C}_1$ &  ${\cal C}_2$ &  ${\cal C}_3$ &  ${\cal C}_4$ &  ${\cal C}_5$ &  ${\cal C}_6$ \\
        \hline
      $1 - \beta^2$  &  $\chi^2 - \zeta^2$ & $\delta^2 - 1$ & $2 \beta \zeta -2\chi$ & $2 \delta + 2 \beta$ & $- 2 \chi \delta - 2 \zeta$ \\ \hline 
    \end{tabular}
    \caption{Relation between the $a(\bm{\psi})_{ij}$ and the ${\cal C}_i$ coefficients.}
    \label{tableac}
\end{table}

Equation~(\ref{HTinverted*}) can be rewritten in the general form~\cite{Polyanin}: 
\begin{equation}
 \sum_{i,j=u,v,\phi}  a_{ij}({\bm{\psi}})  \frac{\partial^2 }{\partial \psi_i \partial \psi_j} \Psi(u,v,\phi) = 
 {\cal F}\Biggl( \bm{\psi};  \frac{\partial }{\partial u}; \frac{\partial }{\partial v};  \frac{\partial }{\partial \phi}  \Biggr)  \Psi(u,v,\phi) , \label{ke}
 \end{equation}
where the $a({\bm{\psi}})_{ij}$ denote functions that may contain linear series and derivatives of the variables $u$, $v$, and $\phi$, even to the second order 
and $\bm{\psi}_i = \{u, v, \phi\}$. Moreover, the right-hand side of this equation may be nonlinear and only the left-hand side is required for its classification. 

At a specific point $\bm{\psi} = \bm{\psi}_0$, the following quadratic form may be assigned to equation (\ref{HTinverted*})
\begin{equation}
    Q =  \sum_{i,j=u,v,\phi}  a_{ij}({\bm{\psi}_0}) \vartheta_i \vartheta_j. \label{Q}
\end{equation}
Applying an appropriate linear non-degenerate transformation
\begin{equation}
    \vartheta_i = \sum_{\kappa = u, v, \phi}\varrho_{i\kappa} \lambda_{\kappa}, \quad \mbox{with} \quad (i = u, v, \phi), \label{Qlinear}
\end{equation}
the quadratic form (\ref{Q}) can be reduced to the canonical form
\begin{equation}
    Q = \sum_{i=u, v, \phi} c_i \lambda^2_i, \label{Qc}
\end{equation}
where the coefficients $c_i$ assume the values $1$, $-1$, and $0$. 
By introducing the new independent variables ${\bm y} = (\eta, \xi, \varphi)$, in accordance with
\begin{equation}
y_i = \sum_{\kappa =u,v,\phi} \varrho_{i \kappa} x_{\kappa}, \quad \mbox{with} \quad x_{\kappa} = (u, v, \phi), \label{yi}
\end{equation}
where the $\varrho_{i \kappa} x_{\kappa}$ are the coefficients of the linear transformation (\ref{Qlinear}), equation (\ref{ke}) is reduced to the canonical form (for the details
see~\cite{Polyanin}):
\begin{equation}
 \sum_{i = \eta, \xi, \varphi} c_i  \frac{\partial^2 }{\partial \varphi^2_i} \Psi(\eta, \xi, \varphi) = 
 {\cal F}\Biggl( \bm{\psi};  \frac{\partial }{\partial \eta};  \frac{\partial }{\partial \xi};  \frac{\partial }{\partial \varphi} \Biggr)  \Psi(\eta, \xi, \varphi) . \label{+ke}
 \end{equation}
\section*{References}

\bibliographystyle{iopart-num.bst} 
\bibliography{main}

\end{document}